\documentclass[11pt]{iopart}

\usepackage{graphicx}

\usepackage{iopams}

\usepackage{amsfonts}
\newcommand{\field}[1]{\mathbb{#1}}
\newcommand{\RR}{\field{R}}

\begin{document}

\title[The electric and magnetic dipole]
{Considerations about the measurement of the magnetic moment and electric dipole moment of the electron}

\author{Mart\'{\i}n Rivas}
\address{Physics Department, The University of the Basque Country,\\ 
Apdo.~644, 48080 Bilbao, Spain}
\ead{martin.rivas@ehu.eus}

\begin{abstract}
The goal of measuring the magnetic moment of the electron, $\bmu$, is to experimentaly determine the gyromagnetic ratio $g$. The factor $g/2$ is computed by precise measurement of two frequencies, the spin precession frequency $\nu_s$, and the cyclotron frequency $\nu_c$, which is defined as $\nu_s/\nu_c=g/2$. These experiments are performed using a single electron confined inside a Penning trap. The existence of the electric dipole moment ${\bi d}_e$, involves the idea of an asymmetric charge distribution along the spin direction such that ${\bi d}_e=d_e{\bi S}/(\hbar/2)$. The energy shift $\Delta U=2{d}_eE_{eff}$ of the interaction of the electric dipole of electrons with a huge effective electric field ${\bi E}_{eff}$, close to the nucleus of heavy neutral atoms or molecules, is calculated by a spin precession measurement and the value $d_e$ is determined. \\
Using a classical model of a spinning electron, which satisfies Dirac's equation when quantized, we determine classically the time-averaged value of the electric and magnetic dipole moments of this electron model when it moves in a Penning trap, with the same fields as in the real experiments, and obtain an estimated value of these dipoles. We compare these results with the experimental data and make some interpretation of the measured dipoles. The conclusion is that experiments do not measure what they purport to measure.

\end{abstract}

\hspace{1.2cm}{\small\bf Keywords:} {Spinning electron; gyromagnetic ratio; magnetic moment; electric dipole moment}

\maketitle

\section{Introduction}
In the analysis of the electron structure, Dirac \cite{Dirac} found two interacting terms, $-{\bi d}{\cdot}{\bi E}-{\bmu}{\cdot}{\bi B}$, which he interpreted as if {\it the electron behave as though it has a magnetic moment $\bmu$ and an electric dipole moment ${\bi d}$. This magnetic moment is assumed in the spinning electron model}, in his words. Dirac disliked the existence of the electric dipole, perhaps because it could be related to some asymmetry of the spherical charge distribution of the electron. In his words, {\it it is doubtful whether the electric moment has a physical meaning}. Two years later, when he published his book \cite{Diracbook}, there was not a single mention of the electric dipole moment.

The predicted magnetic moment $\mu_B=e\hbar/2m$, called Bohr's magneton, is the expected magnetic moment that justifies Zeeman's effect and the hyperfine structure of the hydrogen atom, and it is one of the greatest predictions and successes of Dirac's theory. Preliminary measurements of the magnetic moment showed that its value was different from the predicted value $\mu=(g/2)\mu_B$, where the coefficient $g\neq2$, is called the gyromagnetic ratio, and is considered an intrinsic property of the electron. Since then, theoretical and experimental physicists have also considered the possibility of the existence of the predicted electric dipole moment, and very precise experiments have been designed to measure both properties. 

For the measurement of the magnetic moment, the challenge is to isolate a single electron in a cavity under external electric and magnetic fields and to monitor the excited states to determine this property. In Section {\bf \ref{sec:pening}}, we describe the Penning trap. In Section {\bf\ref{nonrel}}, we describe the motion of a spinless point particle in the cavity to demonstrate the expected motion and resonant frequencies. Because it is a spinless model, we cannot describe the motion of the spin, which is going to be relevant for the determination, at least, of the magnetic moment orientation. The next two sections {\bf\ref{experimentg}} and {\bf\ref{experimentd}} are a description of some experimental setups for the measurements of both dipoles.

Section {\bf\ref{clasicalelectron}} presents a summary of the classical description of a spinning electron, obtained from a general formalism for classical elementary spinning particles \cite{Rivasbook}. The main feature is that the electron is described by the evolution of a single point ${\bi r}$, considered as the center of charge, moving at the speed of light and satisfying a system of fourth-order differential equations. Its trajectory has curvature and torsion; therefore, its classical helical trajectory suggests the so called {\it zitterbewegung} motion. The center of mass is a different point than the center of charge, and it is defined in terms of the point ${\bi r}$ and their derivatives. The center of mass is some average of the position of the evolution of the center of charge. In this section it is also analyzed the motion of the spinning particle in a uniform magnetic field to show that the spin precesses backwards with Larmor's angular velocity $\omega_s=-\omega_c/2$, where $\omega_c=eB/m$, is the cyclotron angular velocity. We end this section with a classical relativistic definition of the electric dipole moment and magnetic moment for a spinning particle, suggested by the previous classical structure. The definition of the magnetic moment of a spinning particle is different than for the spinless point particle. If we accept that this model is a classical description of the electron, then it is the relative motion of the center of charge around the center of mass that produces the magnetic moment. The existence of the electric dipole moment is not related to any asymmetry in the charge distribution. It is related to the separation between the center of mass and the center of charge.

In Section {\bf\ref{DiracPenning}} we analyze the motion of the classical Dirac particle in the Penning trap and obtain an estimate of the time-averaged magnetic moment and electric dipole moment. We end with some conclusions at Section {\bf\ref{conclusions}}. The Appendix {\bf\ref{Append}} describes two Mathematica notebooks to perform some calculations.

\section{The Penning trap. The Geonium}
\label{sec:pening}

\begin{figure}[!hbtp]\centering%
 \includegraphics[width=.4\textwidth]{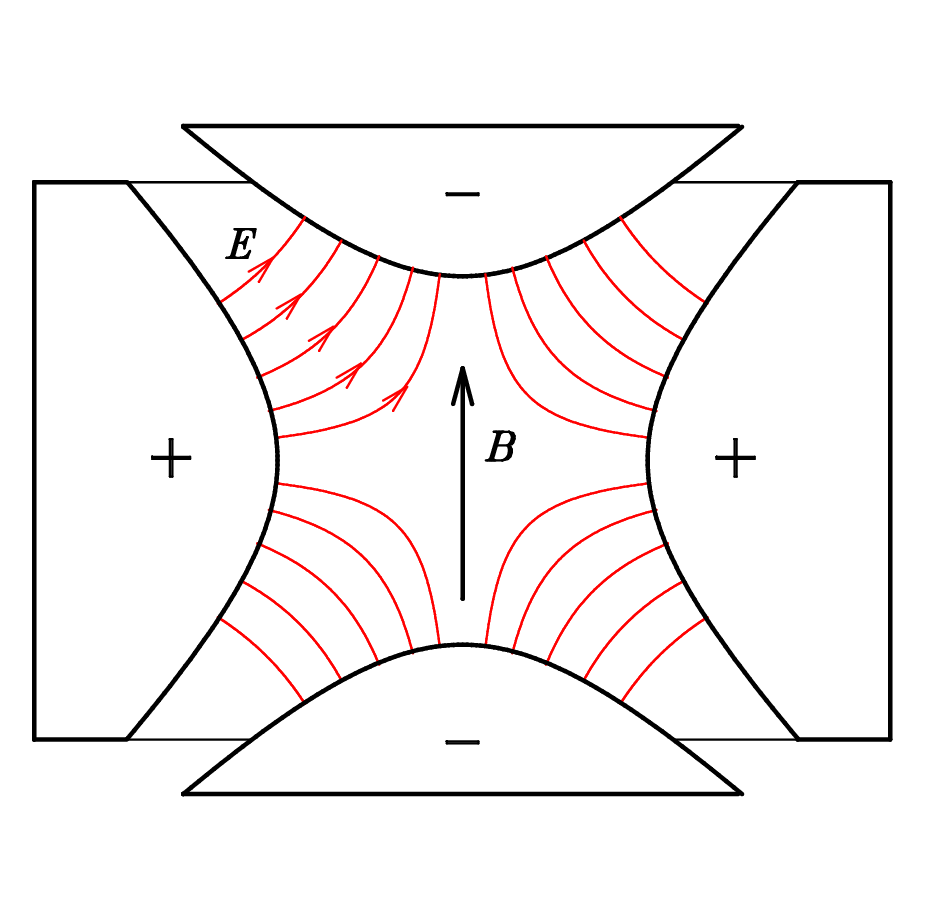}
 \caption{\small{Model of a Penning trap where an electrostatic field between the positively charged lateral wall, and the two negatively charged caps, is established.  Superimposed there is also a uniform axial magnetic field. Red lines correspond to the field lines of the electrostatic field.}}
 \label{fig:Penning1}
\end{figure}

The Penning trap is a cylindrical cavity, like the one depicted in the figure {\bf\ref{fig:Penning1}}. It is bounded by a positively charged one-sheet hyperboloid and two caps, that have the shape of a  negatively charged two-sheet hyperboloid. Inside, a vacuum has been created, it is cooled to a very low temperature and there is also a uniform magnetic field in the axial direction.

If electrons are injected into the cavity, they start to rotate due to the magnetic field. If they have an initial velocity along the magnetic field, when they approach the negatively charged caps they will be repelled and thus the electrons get trapped inside the cavity.
The walls of the cavity are equipotential surfaces of the electrostatic
potential. The electrostatic potential in cylindrical coordinates has the form
\[
V(z,r)=V_0\frac{r^2-2z^2}{4d^2},\quad \nabla^2 V=0.
\]
It satisfies Laplace's equation in the cavity. The surfaces $2z^2-r^2=$ const. are equipotential surfaces. If this constant is positive the origin of the reference system is at the center of the device and the vertex of each cap is at the position $(0,0,\pm z_0)$, the equation of the two-sheeted hyperboloid corresponding to the caps, in Cartesian coordinates, is
\[
z^2=z_0^2+(x^2+y^2)/2.
\]
The surfaces $2z^2-r^2=$ cte $<0$, are also equipotential surfaces. If 
$r_0$ is the radius of the throat of the one-sheeted lateral wall, the equation of this hyperboloid is
\[
z^2=\frac{1}{2}(x^2+y^2-r_0^2).
\]
The equipotential surface $2z^2-r^2=0$, is the cone, centered at the origin, whose
generatrices define the asymptotic directions of the two hyperboloids and of the equipotential surfaces of the electrostatic field.
Therefore,
\[
V(z_0,0)-V(0,r_0)=-\frac{V_0z_0^2}{2d^2}-\frac{V_0r_0^2}{4d^2},
\]
and if we choose for the parameter $d$, the value
\[
d^2=\frac{1}{2}(z_0^2+r_0^2/2),\quad V(z_0,0)-V(0,r_0)=-V_0<0,
\]
then $V_0$, is the difference of the electrostatic potential between the two hyperboloids, being positive the lateral one-sheeted hyperboloid (as shown in the figure {\bf\ref{fig:Penning1}}). 

We take for the vector potential of the uniform magnetic field  ${\bi B}$, 
\[
{\bi A}=\frac{1}{2}{\bi B}\times{\bi r}=\frac{1}{2}Br\widehat{\bi e}_\phi,\quad \nabla\times{\bi A}={\bi B}=B\widehat{\bi e}_z.
\]
The Lagrangian which describes the electron in this cavity is:
\[
{L}={L}_0-eV+e\dot{\bi r}\cdot{\bi A},\quad \dot{\bi r}\cdot{\bi A}=\frac{1}{2}{\bi B}\cdot({\bi r}\times\dot{\bi r}),
\]
where ${\bi r}$ and $\dot{\bi r}$, are, respectively, the position and velocity of the center of charge of the electron in the laboratory reference system, and
${L}_0$ is the free Lagrangian.
In Cartesian coordinates it appears as:
\[
L=L_0-eV_0\frac{2z^2-(x^2+y^2)}{4d^2}+\frac{eB}{2}(x\dot{y}-y\dot{x}),
\]
where the overdot means taking the time derivative.

\section{Non-relativistic spinless point particle in a Penning trap}
\label{nonrel}
\begin{figure}[!hbtp]\centering%
 \includegraphics[width=.5\textwidth]{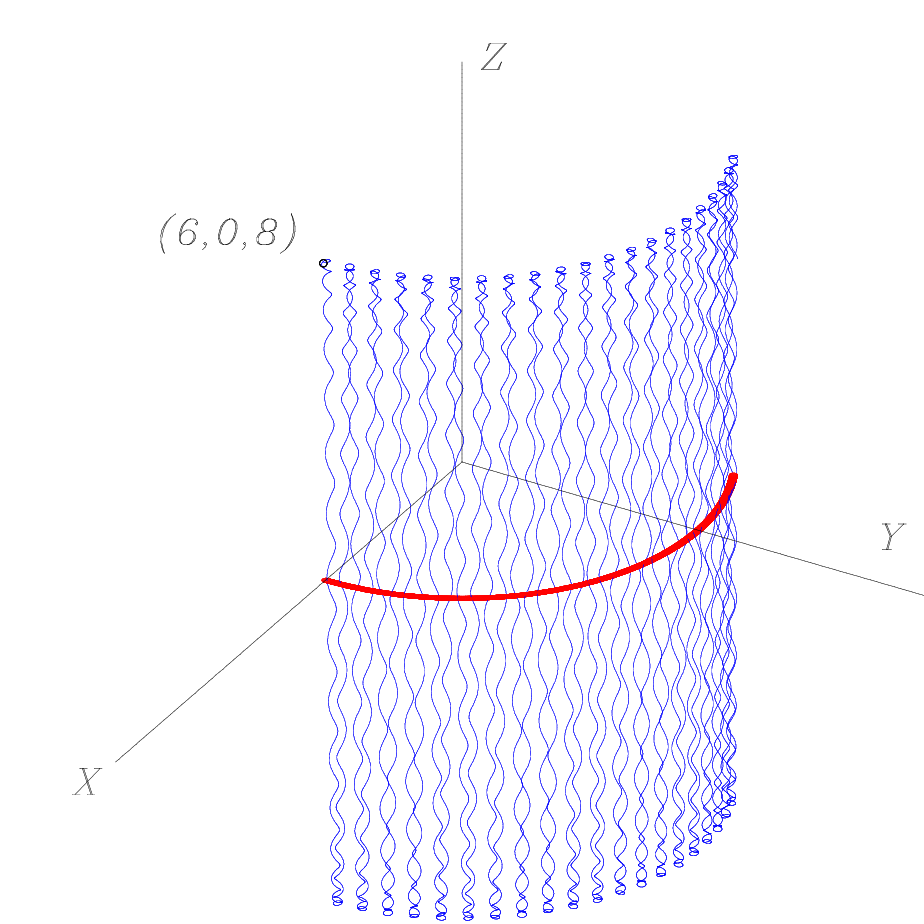}
 \caption{\small{Motion of the spinless point particle (blue), with initial position in $(x_0,y_0,z_0)=(6,0,8)$, initial velocity $(\dot{x}_0,\dot{y}_0,\dot{z}_0)=(0,0.1,0)$, and where
the parameter $a=0.001$. It is also depicted the projection of the motion (red) on the $XOY$ plane.}}
 \label{fig:3DPenning}
\end{figure}

In the non-relativistic framework and for the spinless point particle, the center of charge and center of mass of the electron are the same point and the free Lagrangian $L_0$, is
\[
L_0=\frac{m}{2}\left(\dot{x}^2+\dot{y}^2+\dot{z}^2\right).
\]
The Lagrangian $L$ can be written in two parts, a function of only 
$z$ and $\dot{z}$, and another that depends on the variables $x,y,\dot{x},\dot{y}$,
\[
L=L_1(z,\dot{z})+L_2(x,y,\dot{x},\dot{y}),
\]
where
\[
L_1(z,\dot{z})=\frac{m}{2}\dot{z}^2-\frac{eV_0}{2d^2}z^2,
\]
\[
L_2(x,y,\dot{x},\dot{y})=\frac{m}{2}(\dot{x}^2+\dot{y}^2)+\frac{eV_0}{4d^2}(x^2+y^2)+\frac{eB}{2}(x\dot{y}-y\dot{x}).
\]
The dynamical equation corresponding to the variable $z$ is decoupled with the other two and turns out to be
\[
m\ddot{z}+\frac{eV_0}{d^2}z=0,\quad \ddot{z}+\omega_z^2z=0,\quad \omega_z^2=\frac{eV_0}{md^2}>0,
\]
since for the electron $e<0$ and in the cavity we take $V_0<0$. It represents a vertical harmonic oscillation of constant pulsation $\omega_z$. When quantizing this model, there will be a quantized harmonic energy of value $H_z=\hbar\omega_z(n_z+1/2)$, where $n_z$ represents the number of quanta of this mode.
The variables $x$ and $y$, satisfy the equations
\[
\ddot{x}+\omega_c\dot{y}-\frac{1}{2}\omega_z^2x=0,\quad
\ddot{y}-\omega_c\dot{x}-\frac{1}{2}\omega_z^2y=0,\quad \omega_c=\frac{eB}{m}.
\]
If $\omega_z=0$, we obtain a circular motion perpendicular to the magnetic field, with cyclotron angular velocity $\omega_c$. The presence of the last term, which depends on $\omega_z^2$, is the radial atracting force of the lateral wall. If we use a dimensionless time evolution
parameter $\widetilde{t}=\omega_ct$, the equations of the point particle in the cavity are the linear equations: 
\begin{equation}
\ddot{x}=-\dot{y}+\frac{a}{2}x,\quad \ddot{y}=\dot{x}+\frac{a}{2}y,\quad \ddot{z}=-az,\quad a=\frac{\omega_z^2}{\omega_c^2},
\label{pointparticPenning}
\end{equation}
depending on a single dimensionless parameter $a$, and where the overdot represents the derivative with respect to the dimensionless time $\widetilde{t}$. The exact analytical solution is rather cumbersome in the variables $x$ and $y$. This solution can be obtained with the use of the Mathematica notebook {\tt PointParticleinPenningTrap.nb} quoted in the Appendix {\bf\ref{Append}}. The result is a kind of epicyclic motion of angular velocity $\omega_m\approx \omega_z^2/\omega_c$, superimposed on the cyclotron motion of angular velocity $\omega_c$, and a harmonic motion of pulsation $\omega_z$ in the vertical direction.  Equations (\ref{pointparticPenning}) are also invariant under the scale transformation $\{x,y,z\}\to\lambda\{x,y,z\}$, $\lambda\in\RR$, so that the variables can be considered also dimensionless. A possible trajectory of a spinless electron is that of figure {\bf\ref{fig:3DPenning}}, for $a=0.001$ and $\dot{y}_0=0.1$.

These three frequencies
\[
\omega_m=2\pi\nu_m\simeq\frac{\omega_z^2}{2\omega_c}=\frac{V_0}{2Bd^2},\quad \omega_z=2\pi\nu_z=\sqrt{\frac{eV_0}{md^2}}, \quad \omega_c=2\pi\nu_c=\frac{eB}{m},
\]
are independent of the velocity of the electron, but they depend on the external fields. The magnetronic pulsation $\omega_m$, is independent of the kind of the particle, and only depends on the static fields and the dimensions of the Penning trap. The dimensionless parameter $a$, is:
\begin{equation}
a=\frac{m}{e}\frac{V_0}{B^2d^2}.
\label{parametera}
\end{equation}

A typical cavity, the one built by Gabrielse and Dehmelt \cite{Dehmelt1981}, has the dimensions and fields $d=z_0=r_0/\sqrt{2}=0.335$ cm,
$V_0=10.22$ V, $B=5.872$ T.
With these data, the motion of the electron is a superposition of different harmonic motions,
of pulsations:
\[
\omega_m=7.7543\cdot10^4\,{\rm s}^{-1},\quad \omega_z=4.0021\cdot10^{8}\,{\rm s}^{-1},\quad
\omega_c=1.0327\cdot10^{12}\,{\rm s}^{-1}.
\]
The quanta of these modes are of energies:
\[
\hbar\omega_m=5.104\cdot10^{-11}{\rm eV},\quad
\hbar\omega_z=2.634\cdot10^{-7}{\rm eV},\quad
\hbar\omega_c=6.797\cdot10^{-4}{\rm eV},
\]
and frequencies
\[
\nu_m=12.3414\, {\rm kHz},\quad \nu_z=63.6956\, {\rm MHz},\quad\nu_c=164.37047\, {\rm GHz}.
\]
The value of the parameter $a=1.5016\cdot10^{-7}$. In a Penning trap we get that, in general, $\nu_m\ll\nu_z\ll\nu_c$.

The electrons, trapped in this cavity, make transitions by quanta of the above energies. 
Quantum excited states are represented as $|n_c,n_z,n_m>$, where $n_i$, $i=c,z,m$, are the number of the quanta of each mode. When the energy increases, electrons move faster, the radius increases and some electrons can reach the side walls of the cavity and be absorbed by the metal structure. The energy radiated by the electrons is monitored and every time an electron is absorbed, there is a sudden drop in the radiated energy. When the radiated signal finally corresponds to a single electron, the experimenters conclude that the cavity contains a single electron bound to the earth. The authors give the name {\bf Geonium} to this single electron captured in a Penning trap. The experimental challenge has been to successfully monitor these excited quantum states of the spinless particle. However, in the above analysis of the spinless particle there has been no way to determine the magnetic moment of the electron and its relationship to the spin and with the  measurements of the above three frequencies.

The different order of magnitude of these frequencies $\nu_c$, $\nu_z$ and $\nu_m$, suggests that
the fundamental experimental role will be played by the accurate measurement of the frequency $\nu_c$, and to the next order $\nu_z$.

In the cavity, the ground state of the spinless electron corresponds to $|0,0,0>$. With the above fields, the energy of this state is
\[
H_0=\gamma(v)mc^2=mc^2+\hbar(\omega_c+\omega_z+\omega_m)/2>mc^2,
\]
and the velocity of the electron is 
\[
\gamma(v)=1.000000000665396,\quad v/c=0.00003648, \quad  v=10936.4\,{\rm m/s}.
\]
In a nonrelativistic analysis we obtain the same value
\[
\frac{1}{2}mv^2=\hbar(\omega_c+\omega_z+\omega_m)/2,\quad v=10936.4\,{\rm m/s}.
\]
The electron, even at its lowest energy state, is never at rest. With this velocity in the Penning trap, the cyclotron radius is
\[
R_c=v/\omega_c= 1.05901\cdot10^{-8}\;{\rm m}.
\]
This linear relation means that the cyclotron radius increases with the electron velocity.

\section{Experimental measurement of $g$}
\label{experimentg}
We are going to summarize basic aspects of the experiments of the following three works: The oldest one from 1986, which we will call (W) for Washington \cite{VanDyck1986}, and whose main conclusions were presented in the Letter \cite{VanDyck1987};
the most recent one of 2023, we are going to call (N) for Northwestern \cite{Fan2023}, that makes a comparison of its results with an intermediate one from 2008, called (H) for Harvard \cite{Hanneke2008}. These symbols make reference to the universities where these experiments took place.

The experimental challenge is to isolate a single electron in a Penning trap at very low temperature, interact and excite the electron to produce different transitions and measure the aforementioned frequencies of the different modes.
The aim of these experiments is to accurately measure the cyclotron angular velocity $\omega_c$, and the spin precession angular velocity $\omega_s$, which has not been considered in the previous  analysis of the point particle. 

The hypothesis of these works is that the different energy levels of this single spinning electron in the cavity are quantized according to
\begin{equation}
H=H_s+H_c=\hbar\omega_sm_s+\hbar\omega_c\left(n+\frac{1}{2}\right),
\label{hipotesis}
\end{equation}
where $m_s=\pm1/2$, and $n$ represents the number of excited quanta of the cyclotron motion of  angular velocity $\omega_c=eB/m$. The other modes $\omega_m$ and $\omega_z$, are considered not very relevant. The mode of harmonic frequency $\omega_c$ is quantized with the usual quantization of a one-dimensional harmonic oscillation, but there is no physical justification of the quantization formula ({\ref{hipotesis}}) of the spin precession frequency $\omega_s$. 

We know, as we shall see later in Section \ref{sec:uniformB}, that if a spinning particle in a uniform magnetic field describes a cyclotron trajectory of angular velocity $\omega_c$, the spin precesses backwards with Larmor's angular velocity $-\omega_c/2$. In these experiments, even though they do not have a theoretical interpretation of how the electron spin moves, they assume that the spin precession angular velocity is $g$ times this value, $\omega_s=g\omega_c/2$. In this way $\omega_s/\omega_c=g/2$, and the accurate measurement of these two frequencies will give the value of $g$ and therefore, as they assume, of the magnetic moment:
\[
\frac{g}{2}=\frac{\omega_s}{\omega_c}=1+\frac{\omega_s-\omega_c}{\omega_c}=1+\frac{\omega_a}{\omega_c}, \quad \omega_a=\omega_s-\omega_c.
\]
In (N) they denominate $\nu_a=\omega_a/2\pi$, the anomaly frequency, and the {\it measurement of $\nu_a/\nu_c$ rather than $\nu_s/\nu_c$, significantly reduces the effect of frequency measurement uncertainties}, in their words. For example, they mention,  that in a magnetic field of $B=5.3$ T,  they obtain the following approximate values:
$\nu_c=\omega_c/2\pi\approx$149 GHz, $\nu_a=\omega_a/2\pi\approx$173 MHz. This gives the value
\[
\frac{g}{2}=1.001161073825504,
\]
which is a good aproximation to the expected gyromagnetic ratio. What they do is take a sufficient number of measurements of these frequencies under different magnetic fields and make the corresponding statistical average. In the figure 4 of the cited (N) paper they mention that they have done 11 measurements with different values of the magnetic field $B$, in the range of 3 to 5.5 Teslas.

As a summary of these works the following measurements of $g/2$, have been obtained:
\begin{eqnarray}
1986 \quad (W)\quad {g}/{2}&=&1{\rm.}001\,159\,652\,200\,(40),\cr
2008 \quad (H)\quad {g}/{2}&=&1{\rm.}001\,159\,652\,180\,73\,(28),\cr
2023 \quad (N)\quad {g}/{2}&=&1{\rm.}001\,159\,652\,180\,59\,(13).\nonumber
\end{eqnarray}
In this last measurement (N), they ensure that it is 2.2 times more accurate than in (H), from the statistical point of view.

In the Van Dyck et al. work (W) the Penning trap is something different as the one described above, with an electrostatic potential 
\[
V(z,r)=V_0\frac{r^2-2z^2}{4d^2}+V_0C_4\frac{8z^4-24r^2z^2+3r^4}{16d^4},
\]
where the coefficient $C_4$ depends on the geometry of the cavity. It is cooled to 4 K and a magnetic field of 1.8 T is used.
In the other two works (N) and (H) the authors mention that the Penning trap is cylindrical.
In (N) the cavity is cooled to 50 mK, with a surrounding solenoid cooled to 4.2 K and the magnetic field is in the range of 3 to 5.5 Teslas.

However, the main experimental hypothesis ({\ref{hipotesis}}) and that the spin precession frequency is exactly $\omega_s/\omega_c=g/2$, are not theoretically justified. In our classical analysis in a uniform magnetic field the spin always precesses backwards with  Larmor's angular velocity $\omega_s=-\omega_c/2$, as we will see in Section {\bf\ref{clasicalelectron}} and in the figure {\bf\ref{fig:Rc5}}.

\section{Experimental measurement of the Electric dipole moment}
\label{experimentd}
\begin{figure}[!hbtp]\centering%
\hspace{1cm}\includegraphics[width=14cm]{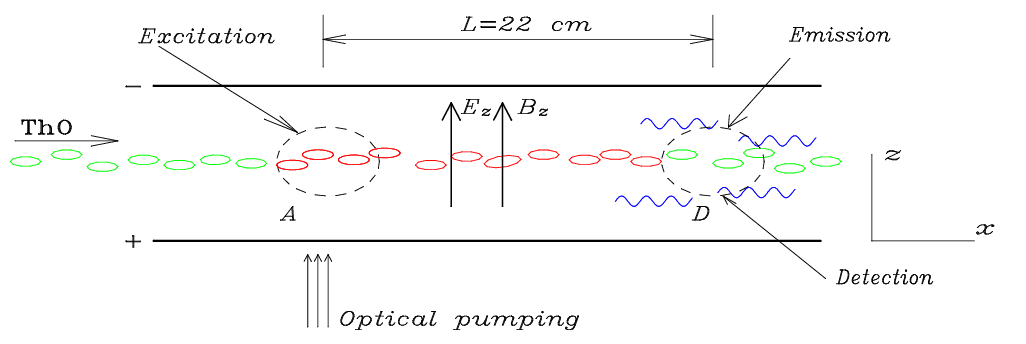}\caption{Evolution of the ThO molecules inside the cavity. In the region $A$ where some molecules are excited (green to red) the corresponding excited electrons reverse the orientation of their electric dipole, absorbing some photons from the optical pumping. Later, the molecule makes a spontaneous emission in region $D$, where the detectors are located, and the energy of the emitted photons is measured.}
\label{fig:MeasureDipol}
\end{figure}

We are going to analyze a recent experiment \cite{ACME}, to measure the electric dipole moment of the electron. The experimental setup consists of sending a beam of neutral heavy atoms or molecules, in this case thorium monoxide (ThO), between two charged plates, parallel to the XOY plane, separated 25 mm. There is a uniform electric field $E_z$, and a uniform magnetic field $B_z$,
orthogonal to the plates, as depicted in figure {\bf\ref{fig:MeasureDipol}}. The use of heavy atoms is because the effective electric field $E_{eff}$, at points very close to the nucleus is around $10^6$ times stronger than any electric field produced at the laboratory. If the electrons have an electric dipole ${\bi d}_e$, in their ground state they will be oriented along the efective electric field, so that the energy of that state is $-d_eE_{eff}$. If we excite the molecules by transverse optical pumping in area $A$, some electrons will jump to the excited state with an inverted orientation of their electric dipole, after absorbing an energy of value $\Delta U=2d_eE_{eff}$. This energy will be determined in the detection area $D$, by measuring the energy of the photons emitted in this spontaneous emission. 

The main theoretical hypothesis is that the electric dipole has the same orientation as the spin ${\bi d}_e=d_e{\bi S}/(\hbar/2)$, and if the molecules interact with the external field $B_z$, the  precession of the spin will also be the precession of the electric dipole. The magnetic moment of the excited electron is assumed to be $(g/2)\mu_B$, and during the time of flight of the molecules interacts  with the external magnetic field $B_z$, and the spin will precesses through an angle $\phi$, in the $XOY$ plane, which is the same angle of precession as the electric dipole moment. This precession angle is the one that will be measured at the readout section $D$ of the experiment, where the fluorescence detectors are placed.

The experimental hypothesis is that this precession angle of the spin is
\begin{equation} 
\phi\approx \left(-g\mu_B B_z-d_eE_{eff}\right)\tau/\hbar,
\label{anglefi}
\end{equation}
where $\tau\approx1.1$ ms, is the time of flight of the molecules between the excitation area $A$ and detection area $D$, which are separated by a distance $L\approx 22$ cm. If $\phi$ and $\Delta U$, are accurately measured, they determine the values of $d_e$ and $E_{eff}$.

The experimental result suggests that the upper bound for the electric dipole is
\[
d_e<8.7\cdot10^{-29}\,e\,{\rm cm}.
\]

\section{Classical Dirac particle}
\label{clasicalelectron}

The model of a classical elementary spinning particle that we are going to use in this work, has been obtained in a general formalism \cite{Rivasbook}, that is based on the following three fundamental principles: Restricted Relativity Principle, Variational Principle and Atomic Principle. 

\begin{figure}[!hbtp]\centering%
 \includegraphics[width=.5\textwidth]{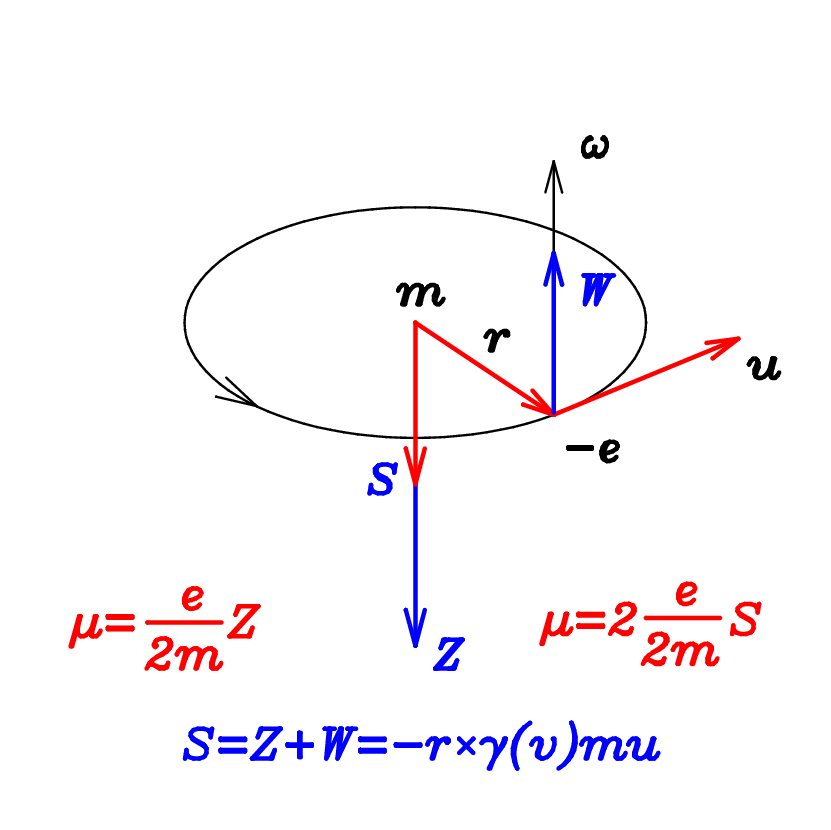}
 \caption{\small{This model represents the circular motion, at the speed of light, of the center
of charge of the electron in the center of mass frame. The center of mass is always a different point that the center of charge.
The radius of this motion is $R_0=\hbar/2mc$, in this frame.
The angular velocity is $\omega_0=2mc^2/\hbar$. This frequency, decreases when the center of mass moves. The local clock slows down when moving. The spin has two contributions: one ${\bi Z}$ from the orbital motion of the CC around the CM (Zitterbewegung) and another ${\bi W}$ in the opposite direction, related to the rotation with angular velocity $\bomega$ of a comoving Cartesian frame attached to the CC. The magnetic moment is related to the motion of the CC, and therefore to the zitterbewegung contribution to the spin, ${\bi Z}$, with a normal relation ${\bmu}=e{\bi Z}/2m$. It is when we expressed the magnetic moment in terms of the total spin ${\bi S}$, is when we the concept of gyromagnetic ratio $g=2$ \cite{Rivasg=2} is obtained.}
}
 \label{fig:elecCM}
\end{figure}

The definition of a classical elementary particle lies on the Atomic principle \cite{atomic}. The idea is that an elementary particle does not have internal excited states in the sense that any interaction, if it does not annihilate the particle, does not modify its internal structure. If an inertial observer describes any state of the elementary particle in a Lagrangian description through a set of variables $(x_1,\ldots,x_n)$ and the dynamics changes this state to  another $(y_1,\ldots,y_n)$, if the internal structure has not been modified, then it is possible to find another inertial observer who describes this new state of the particle with exactly the same values of the variables as in the previous state. This means that there must exist a Poincar\'e transformation $g$, between both inertial observers such that the above values are transformed into each other: 
\[
(y_1,\ldots,y_n)=g\,(x_1,\ldots,x_n).
\]
This relation must be valid for any pair of states of any elementary particle.
The Atomic Principle restricts the classical variables of the variational description of an elementary particle, to belong to a homogeneous space of the Poincar\'e group \cite{Rivasbook}. Then the initial and final states of the Lagrangian description are described at most by as many variables as those of any Poincar\'e group element, and with the same geometrical or physical meaning like the group parameters. The initial state is characterized by at most ten variables $x_i\equiv(t,{\bi r},{\bi u},\balpha)$, that are interpreted as the time $t$, the position of a single point ${\bi r}$, the velocity of this point ${\bi u}$,
and finally the orientation $\balpha\in SO(3)$ of a comoving Cartesian frame attached to the point ${\bi r}$. The same variables, with different values, for the final point $x_f$ of the Lagrangian evolution.
Among the models of spinning particles this formalism predicts, the only one that satisfies Dirac's equation when quantized \cite{RivasDirac}, is that model in which the point ${\bi r}$, is moving at the speed of light $u=c$. Since the Lagrangian also depends  on the next order time derivative of the above boundary variables, the Lagrangian must depend on the acceleration of the point ${\bi a}$, and on the angular velocity $\bomega$. The point ${\bi r}$, satisfies therefore  a system of fourth-order differential equations and, being the only point where the external fields are defined, we interpret the point ${\bi r}$, as the location of the center of charge (CC) of the particle.

We must remember that according to Frenet-Serret formalism, the family of continuous and differentiable trajectories of points with continuous and differentiable curvature and torsion
in three-dimensional space, satisfy a system of fourth-order differential equations.

This model is depicted in figure {\bf{\ref{fig:elecCM}}} for the center of mass observer. It has a center of charge (CC), ${\bi r}$, that moves at the speed of light $c$, around the center of mass (CM), ${\bi q}$, that is at rest in this frame, and that is a different point than the CC. In the free motion the CC describes a circle of radius $R_0$, contained in a plane orthogonal to the spin. We will call this model from now on a {\bf classical Dirac particle}.  

The position ${\bi r}$ of the CC, satisfies a system of fourth-order differential equations
\cite{RivasDynamics}, that can be separated into a system of second-order differential equations for the CC and CM positions, where the CM position ${\bi q}$, is defined in terms of the motion of the CC, ${\bi r}$, by 
\begin{equation}
{\bi q}={\bi r}+\left(\frac{c^2-{\bi v}\cdot{\bi u}}{(d^2{\bi r}/dt^2)^2}\right)\frac{d^2{\bi r}}{dt^2},
\label{defposCM}
\end{equation}
where
${\bi v}={d{\bi q}}/{dt}$, and ${\bi u}={d{\bi r}}/{dt}$ and $v<c$ and $u=c$.
The energy $H$, and the linear momentum ${\bi p}$, are written, as in the case of the point particle model, in terms of the center of mass velocity ${\bi v}$, as:
\begin{equation}
H=\gamma(v)mc^2,\quad {\bi p}=\frac{H}{c^2}{\bi v}=\gamma(v)m{\bi v}.
\label{defHyp}
\end{equation}

The relativistic differential equations in the presence of an external electromagnetic field ${\bi E}(t,{\bi r})$ and ${\bi B}(t,{\bi r})$, defined at the CC position ${\bi r}$, and in any arbitrary inertial reference frame are \cite{RivasDynamics}: 
\begin{eqnarray}
\frac{d^2{\bi q}}{dt^2}&=&\frac{e}{m\gamma(v)}\left[{\bi E}+{\bi u}\times{\bi B}-\frac{1}{c^2}{\bi v}\left(\left[{\bi E}
+{\bi u}\times{\bi B}\right]\cdot{\bi v}\right)\right],\label{eq:d2qdt2}\\
\frac{d^2{\bi r}}{dt^2}&=&\frac{c^2-{\bi v}\cdot{\bi u}}{({\bi q}-{\bi r})^2}\left({\bi q}-{\bi r}\right),\label{eq:d2rdt2}
 \end{eqnarray}
where
$\gamma(v)=(1-{\bi v}^2/c^2)^{-1/2}$, and the constraint $|{\bi u}|=c$. The Lagrangian which gives rise to these dynamical equations is
\[
L=L_0({\bi u},{\bi a},\bomega)+L_{em}(t,{\bi r},{\bi u}),\quad L_{em}=-eA_0(t,{\bi r})+e{\bi u}\cdot{\bi A}(t,{\bi r})
\]
where $L_0$ is the free Lagrangian, $e$ is the electric charge of the particle and $A_0$ and ${\bi A}$, are, respectively the scalar and vector potentials defined at the CC of the particle and ${\bi u}$ is the CC velocity.

The free Lagrangian $L_0({\bi u},{\bi a},\bomega)$, is translation invariant and therefore it is a function of the velocity, acceleration and angular velocity of the CC. The fourth-order Euler-Lagrange equations of the position variables are\footnote{\footnotesize{We represent 3D vector observables in boldface. Expressions like ${\bi a}=\partial L/\partial {\bi b}$, have to be interpreted throughout this work as $a_i=\partial L/\partial b_i$, $i=1,2,3$.}}
\[
\frac{\partial L_0}{\partial{\bi r}}-\frac{d}{dt}\left(\frac{\partial L_0}{\partial{\bi u}}\right)+
\frac{d^2}{dt^2}\left(\frac{\partial L_0}{\partial{\bi a}}\right)+\frac{\partial L_{em}}{\partial{\bi r}}
-\frac{d}{dt}\left(\frac{\partial L_{em}}{\partial{\bi u}}\right)=0.
\]
Since $L_0$ is independent of ${\bi r}$, the above dynamical equation reads:
\[
-\frac{d}{dt}\left[\frac{\partial L_0}{\partial{\bi u}}-
\frac{d}{dt}\left(\frac{\partial L_0}{\partial{\bi a}}\right)\right]-e(\nabla A_0+\partial{\bi A}/\partial t)+e{\bi u}\times(\nabla\times{\bi A})=0,
\]
and the term between the squared brackets represents the mechanical linear momentum of the particle ${\bi p}$.
This equation is
\[
\frac{d{\bi p}}{dt}=e\left({\bi E}+{\bi u}\times{\bi B}\right),
\]
and corresponds to equation (\ref{eq:d2qdt2}) 
where ${\bi u}$ is the velocity of the CC, which appears in the magnetic force. From ${\bi p}$  defined in (\ref{defHyp}) its time derivative is expressed in terms of $d^2{\bi q}/dt^2$, and writing on the left hand side this derivative we get (\ref{eq:d2qdt2}).
Equation (\ref{eq:d2rdt2}) is equation (\ref{defposCM}) when writing on the left hand side the acceleration of the CC. Taking the second order derivative of ${\bi q}$ in (\ref{defposCM}) and replacing it by (\ref{eq:d2qdt2})  the fourth-order system of differential equations for the CC position ${\bi r}$ is recovered \cite{RivasDynamics}, which is where the external fields ${\bi E}(t,{\bi r})$ and ${\bi B}(t,{\bi r})$, are defined. A more detailed description of how the dynamical equations and the definition of the different observables are obtained can be found in \cite{ClassicalDirac1}.

The point particle free Lagrangian is a function $L_0({\bi u})$ of the velocity variable of the
point ${\bi r}$ and is independent of the time $t$ and position ${\bi r}$. The angular momentum with respect to the origin is obtained by applying the rules of Noether's theorem  and gives ${\bi J}={\bi r}\times{\bi p}$, where the linear momentum is obtained from the Lagrangian as ${\bi p}=\partial L_0/\partial{\bi u}$. 
For the spinning particle, the free Lagrangian is a function $L_0({\bi u},{\bi a},\bomega)$. Nother's theorem defines the angular momentum with respect to the origin as ${\bi J}={\bi r}\times{\bi p}+{\bi u}\times{\bi U}+{\bi W}$, where ${\bi U}=\partial L_0/\partial{\bi a}$, ${\bi W}=\partial L_0/\partial{\bomega}$ and ${\bi p}=\partial L_0/\partial{\bi u}-d{\bi U}/dt$. Therefore ${\bi J}={\bi r}\times{\bi p}+{\bi S}$, where the function ${\bi S}={\bi u}\times{\bi U}+{\bi W}$, represents the angular momentum with respect to the point ${\bi r}$. It is the spin with respect to the CC. The additional variables ${\bi a}$ and $\bomega$ in the Lagrangian, that define the observables ${\bi U}$ and ${\bi W}$, provide the structure of the spin when we compare it with the point particle.

Since this model has two characteristic points, the formalism defines the angular momentum (spin) of the particle with respect to both points ${\bi r}$ and ${\bi q}$. It is found that the spin with respect to the CC, ${\bi r}$, is:
\begin{equation}
{\bi S}=-\gamma(v)m({\bi r}-{\bi q})\times{\bi u},
\label{eq:spinCC}
\end{equation}
while the spin with respect to the CM, ${\bi q}$, is defined by:
\begin{equation}
{\bi S}_{CM}={\bi S}+{(\bi r}-{\bi q})\times{\bi p}=-\gamma(v)m({\bi r}-{\bi q})\times({\bi u}-{\bi v}).
\label{eq:spinCM}
\end{equation}
The spins (\ref{eq:spinCC}) and (\ref{eq:spinCM}) are finally expressed in terms of the instantaneous separation between both centers and of the velocities of both points, while it is only the velocity of the CM ${\bi v}$, which appears in the definition of $H$ and ${\bi p}$, (\ref{defHyp}).
 
For the center of mass observer ${\bi q}={\bi v}=0$, and both spins take the same value in this frame. According to the figure {\bf{4}}, if $R_0=|{\bi r}-{\bi q}|$ is the separation between the CC and CM, then the constant value of the spin $S=\hbar/2$,
\[
\frac{\hbar}{2}=mcR_0, \quad \Longrightarrow \quad R_0=\frac{\hbar}{2mc}=1.93\cdot10^{-13}\,{\rm m},
\]
and this separation is half Compton's wavelength, as was mentioned before.

The total angular momentum ${\bi J}$ of the Dirac particle with respect to the origin of any arbitrary inertial frame, can be written in two ways as:
\[
{\bi J}={\bi S}+{\bi r}\times{\bi p}={\bi S}_{CM}+{\bi q}\times{\bi p}.
\]
Both spins satisfy different dynamical equations. If the particle is free, ${\bi v}$ is constant and $d{\bi J}/dt=0$, and $d{\bi p}/dt=0$, leads to
\[
\frac{d{\bi S}}{dt}={\bi p}\times{\bi u},\quad \frac{d{\bi S}_{CM}}{dt}=0.
\]
The CM spin ${\bi S}_{CM}$, is conserved while the spin with respect to the CC, ${\bi S}$, satisfies the same dynamical equation as Dirac's spin operator in the quantum case. Its time derivative is always orthogonal to the direction of the linear momentum. We must remark that in the quantum Dirac's analysis, the only variable that defines the position of the electron ${\bi r}$, is contained in Dirac's spinor $\psi(t,{\bi r})$,  it is moving at the speed $c$ and it is also the point where the external electromagnetic field $A_\mu(t,{\bi r})$ is defined. Clearly, the spin with respect to the CC, ${\bi S}$, represents the classical equivalent of Dirac's spin operator.

The spin ${\bi S}$ has two parts: one ${\bi Z}={\bi u}\times{\bi U}$, associated to this relative internal motion and to the dependence of the Lagrangian on the acceleration and another ${\bi W}$, in the opposite direction, related to the dependence of the Lagrangian on the angular velocity of a local Cartesian frame attached to the motion of the center of charge. This frame is not depicted in the figure. The magnetic moment of the electron is produced by the motion of the charge (zitterbewegung) and is related to the orbital part ${\bi Z}$ of the angular momentum by ${\bmu}=e{\bi Z}/2m$. When the magnetic moment is expressed in terms of the total spin ${\bi S}$, which is half the orbital part ${\bi Z}$, is when the concept of gyromagnetic ratio $g=2$, \cite{Rivasg=2} is obtained.

The energy can also be written in the form:
\begin{equation}
H={\bi p}\cdot{\bi u}+\frac{1}{c^2}{\bi S}\cdot\left(\frac{d{\bi u}}{dt}\times{\bi u}\right),
\label{eq:DiracH}
\end{equation}
which contains two terms: the first term, proportional to the linear momentum, represents the translational energy, while the second, proportional to the spin and to the motion of the CC (zitterbewegung), represents the rotational energy which never vanishes. If we introduce in (\ref{eq:DiracH}) the expressions of ${\bi p}$ and ${\bi S}$, given in (\ref{defHyp}) and (\ref{eq:spinCC}), respectively, we get $H=\gamma(v)mc^2$. It is this linear expression in terms of $H$ and ${\bi p}$, which will give rise to Dirac's Hamiltonian when quantizing the model \cite{RivasDirac}. In the quantum description $H=i\hbar\partial/\partial t$ and ${\bi p}=-i\hbar\nabla$ and (\ref{eq:DiracH}) is transformed into Dirac equation.

\subsection{The angular velocity}
\label{angular}

The angular velocity corresponds to the rotation of the comoving Cartesian frame linked to the point ${\bi r}$. But this frame is completely arbitrary. Once the CC is moving, its dynamics defines the acceleration ${\bi a}$, orthogonal to the velocity ${\bi u}$, and therefore
these two orthogonal vectors with the vector ${\bi u}\times{\bi a}$ define an orthogonal comoving frame (the Frenet-Serret frame). Since ${\bi u}$ is a vector of constant absolute value, its time derivative the acceleration ${\bi a}$, is orthogonal to it and it can be written as
\[
{\bi a}={\bomega}\times{\bi u}.
\]
The cross product of this expression with ${\bi u}$ gives:
\[
{\bi a}\times{\bi u}={\bi u}({\bomega}\cdot{\bi u})-c^2\bomega.
\]
The first term is the component of the angular velocity along the velocity ${\bi u}$, $\bomega_u$ times $c^2$, so that $\bomega-\bomega_u=\bomega_p=({\bi u}\times{\bi a})/c^2$. Then the perpendicular component of the angular velocity $\bomega_p$ to the plane defined by the vectors ${\bi u}$ and ${\bi a}$, can be expressed as
\begin{equation}
\bomega_p=\frac{1}{c^2}({\bi u}\times{\bi a})=\frac{1}{c^2}\left(\frac{d{\bi r}}{dt}\times{\frac{d^2{\bi r}}{dt^2}}\right).
\label{omegaperp}
\end{equation}
This perpendicular component has the opposite direction than the CC spin since from (\ref{eq:spinCC}) and (\ref{eq:d2rdt2}), $({\bi r}-{\bi q})\simeq -{\bi a}$. In general the CM spin and the CC spin are not along the angular velocity $\bomega$. This perpendicular component is responsible for the curvature of the trajectory, while the $\bomega_u$ is related to the torsion of the trajectory. In the center of mass frame the trajectory of the point ${\bi r}$ is flat, there is no torsion and thus $\bomega_u=0$.

The component $\bomega_u$ produces the torsion of the trajectory and depends on the third orden derivative of the vector ${\bi r}$. This component of the angular velocity is
\begin{equation}
\bomega_u=\frac{1}{c^2a^2}\left(({\bi u}\times{\bi a})\cdot\frac{d{\bi a}}{dt}\right){\bi u}=\frac{1}{c^2a^2}\left[\left(\frac{d{\bi r}}{dt}\times{\frac{d^2{\bi r}}{dt^2}}\right)\cdot\frac{d^3{\bi r}}{dt^3}\right]\frac{d{\bi r}}{dt}.
\label{omegau}
\end{equation}

The evolution of the angular velocity is completely determined by the motion of the CC. The evolution of the CC determines both the trajectories of the CC and CM and the rotation of the body frame (Frenet-Serret frame) attached to the point ${\bi r}$.

We can find an alternative expression for the angular velocity $\bomega_u$ if we use the dynamical equation (\ref{eq:d2rdt2}). Taking the next order time derivative we have:
\[
\frac{d^3{\bi r}}{dt^3}=\frac{(-{\bi a}_{CM}\cdot{\bi u}-{\bi v}\cdot{\bi a})({\bi q}-{\bi r})^2-2({\bi q}-{\bi r})\cdot({\bi v}-{\bi u})(c^2-{\bi v}\cdot{\bi u})}{({\bi q}-{\bi r})^4}\,({\bi q}-{\bi r})+
\]
\[
+\frac{c^2-{\bi v}\cdot{\bi u}}{({\bi q}-{\bi r})^2}\,({\bi v}-{\bi u}).
\]
The term in squared brackets in (\ref{omegau})
\[
\left(\frac{d{\bi r}}{dt}\times{\frac{d^2{\bi r}}{dt^2}}\right)\cdot\frac{d^3{\bi r}}{dt^3}=\left(\frac{d^2{\bi r}}{dt^2}\times{\frac{d^3{\bi r}}{dt^3}}\right)\cdot\frac{d{\bi r}}{dt}
=\left({\bi a}\times{\frac{d^3{\bi r}}{dt^3}}\right)\cdot{\bi u},
\]
and since from (\ref{eq:d2rdt2}) ${\bi a}\simeq ({\bi q}-{\bi r})$, the last term in this expression is 
\[
\frac{c^2-{\bi v}\cdot{\bi u}}{({\bi q}-{\bi r})^2}\,({\bi a}\times({\bi v}-{\bi u}))\cdot{\bi u}.
\]
Similarly the term
\[
({\bi a}\times({\bi v}-{\bi u}))\cdot{\bi u}=(({\bi v}-{\bi u})\times{\bi u})\cdot{\bi a}=({\bi v}\times{\bi u})\cdot{\bi a}=({\bi u}\times{\bi a})\cdot{\bi v}.
\]
Collecting terms, instead of the expression (\ref{omegau}) we obtain
\begin{equation}
\bomega_u=\frac{1}{c^2a^2}\frac{c^2-{\bi v}\cdot{\bi u}}{({\bi q}-{\bi r})^2}\left[({\bi u}\times{\bi a})\cdot{\bi v}\right]{\bi u}=\left[\frac{\bomega_p\cdot{\bi v}}{c^2-{\bi v}\cdot{\bi u}}\right]{\bi u}.
\label{omegaunew}
\end{equation}
If the CM velocity ${\bi v}=0$, or along to the zitterbewegung plane, $\bomega_p\cdot{\bi v}=0$ and thus $\bomega_u=0$, the CC trajectory is a flat curve and there is no torsion. If $\bomega_p$ and ${\bi v}$ form an angle $\alpha<\pi/2$, $\bomega_u$ has the direction of ${\bi u}$ and the opposite direction if $\alpha>\pi/2$. 

The evolution of the angular velocity is determined by the motion of the CC.
In this way, the above rotational part of the Hamiltonian (\ref{eq:DiracH}) is just $-{\bi S}\cdot{\bomega}>0$,
because the spin has the direction opposite to the angular velocity of this frame, and never vanishes. The classical equivalent to Dirac's Hamiltonian (\ref{eq:DiracH}) can also be written as $H={\bi p}\cdot{\bi u}-{\bi S}\cdot{\bomega}$, where the translation and rotation energies are more evident.

In the Lagrangian description we started with a six-degree-of-freedom mechanical system. Three represent the position of a point ${\bi r}$, and another three angles $\balpha\in SO(3)$, that describe the orientation of a comoving frame. The dynamics has established a constraint such that the degrees of freedom $\balpha$ are finally related to the motion of the point ${\bi r}$, which satisfies a system of fourth-order differential equations. This point and their derivatives, completely determine the definition of any other observable. The electron description has been reduced to the evolution of a single point, the center of charge.

\subsection{Planck and de Broglie hypothesis}
\label{PlanckDB}
The manifold spanned by the variables $(t,{\bi r},{\bi u},\balpha)$, where $t\in\RR$, ${\bi r}\in\RR^3$, ${\bi u}\in\RR^3$ with $u<c$,  and $\balpha\in SO(3)$, is the Poincar\'e group manifold. The manifold with the constraint $u=c$, is also a homogeneous space of the Poincar\'e group. According to the Atomic Principle it can represent the boundary variables manifold of the Lagrangian description of a classical elementary particle. In this manifold two different kinds of elementary particles can be described \cite{RivasJMP}. One possibility is that the velocity ${\bi u}$ is a constant vector and the point moves along a straight line with constant velocity $c$, and the comoving frame rotates with angular velocity $\bomega$ in the same direction, pointing forward or backwards. The acceleration vanishes and the spin is related to the ${\bi W}$ part. This is the classical description of a photon. The free Lagrangian of a photon $L_0({\bi u},\bomega)$ is 
\[
L_0=\epsilon\frac{S}{c}{\bi u}\cdot\bomega,
\]
where $ \epsilon=\pm1$, represents the helicity.
The spin ${\bi S}=\partial L_0/\partial\bomega=\epsilon S{\bi u}/c$, and is not transversal. The linear momentum ${\bi p}=\partial L_0/\partial{\bi u}=\epsilon S \bomega/c$ and also has the same direction than the velocity ${\bi u}$. All four vectors, ${\bi p}$, ${\bi u}$, ${\bi S}$ and $\bomega$ are collinear vectors. The energy of the photon has two expressions $H={\bi p}\cdot{\bi u}=pc={\bi S}\cdot{\bomega}=S\omega$. Translation and rotation energies are the same. If $S=\hbar$ then $H=\hbar\omega=h\nu$, which is Planck's hypothesis concerning the quanta of the electromagnetic field. The photon is a boson.

The other possibility for $u=c$, is that ${\bi u}\cdot d{\bi u}/dt=0$, and gives rise to the Dirac particle described in the figure {\bf\ref{fig:elecCM}}.  It is a system of six degrees of freedom, ${\bi r}$ and $\balpha$. If we analyze this particle for the center of mass observer,  and we choose the reference frame in such a way that the angular velocity (and the spin) is along $OZ$ axis, it is reduced to a system of three degrees of freedom: the $x$ and $y$ coordinates of the point on the $XOY$ plane and the phase $\alpha$ of the comoving Cartesian frame attached to the point. But the dynamics produces the constraint that the comoving frame can be the Frenet-Serret frame and the phase $\alpha$ is exactly the phase of the rotational circular motion of the point ${\bi r}$. But if the point describes a circle of radius $R_0$, given the coordinate $x$, the coordinate $y$ is determined so that we are left with a system of just one degree of freedom for the center of mass observer. This degree of freedom $x$, describes a one-dimensional harmonic oscillator of pulsation $\omega_0$ in this frame. According to the Atomic Principle it has no excited states and therefore it is a one-dimensional harmonic oscillator in its ground state. The energy of this particle at rest (${\bi p}=0$) is $H=mc^2=-{\bi S}\cdot{\bomega}=S\omega_0=\hbar\omega_0/2$, i.e., the ground energy of the one-dimensional harmonic oscillator, which gives the value $S=\hbar/2$. This object is a fermion. The internal frequency of this internal rotation motion $\nu_0=2mc^2/h$, is not the postulated de Broglie's frequency but the frequency predicted by Dirac, twice de Broglie's frequency. It seems that De Broglie's hypothesis is wrong because the electron is not a boson, it is a fermion. In this formalism the Dirac particle has been reduced in the center of mass frame to a {\bf one-dimensional harmonic oscillator} with no excited states. 

\subsection{Natural units}
\label{Natural}

The system of differential equations (\ref{eq:d2qdt2}) and (\ref{eq:d2rdt2}) can be rewritten in terms of dimensionless variables, once the expressions of the fields are known.
If we replace ${\bi u}=c\widetilde{\bi u}$,
${\bi v}=c\widetilde{\bi v}$, ${\bi r}=2R_0\widetilde{\bi r}$,  ${\bi q}=2R_0\widetilde{\bi q}$ and $t=\tau_0\widetilde{t}$, where $R_0=\hbar/2mc$, is the separation between the CC and CM of the Dirac particle at rest and $\tau_0=2R_0/c$,
the equation (\ref{eq:d2rdt2}) becomes
\[
\frac{d^2{\bi r}}{dt^2}=\frac{c^2}{2R_0}\frac{d^2\widetilde{\bi r}}{d\widetilde{t}^2}=\frac{c^2}{2R_0}
\frac{1-{\widetilde{\bi v}}\cdot{\widetilde{\bi u}}}{({\widetilde{\bi q}}-{\widetilde{\bi r}})^2}({\widetilde{\bi q}}-{\widetilde{\bi r}})
\]
the coefficients cancel and the equation remains of the same form as in (\ref{eq:d2rdt2}) in terms of dimensionless variables and $c=1$.
For the other equation we need the explicit form of the fields. We shall analyze this in different examples.

We have defined two natural units: the speed of light $c$ and the natural unit of length $2R_0=\hbar/mc$, where $R_0$ represents the separation between the CC and the CM. This defines a natural unit of time $\tau_0=2R_0/c$, as the time taken by a light ray to cover the distance $2R_0$. The time of a complete trajectory of the CC around the CM at the center of mass frame is $\pi$. All boundary variables are dimensionless.
 
From the mechanical point of view we need an extra natural unit of mass. The usual assumption is to define as the natural unit of action such that Planck's constant $\hbar=1$. In this way since
\[
2R_0=\frac{\hbar}{mc}=1,\quad\Rightarrow\quad m=1\;{\rm n.u.},
\]
therefore the natural unit of mass is that unit where the mass of the electron is $1$. 
The energy of the particle is expressed in terms of the center of mass velocity as $H(v)=\gamma(v)mc^2$ which in natural units for the electron $H(v)=\gamma(v)$. The linear momentum is expressed as ${\bi p}(v)=\gamma(v){\bi v}$. The value of the spin in the center of mass frame $S=\hbar/2=1/2$ in natural units. The general expressions of both spins for the electron in natural units are
\[
{\bi S}=-{\gamma(v)}({\bi r}-{\bi q})\times{\bi u},
\]
\[
{\bi S}_{CM}=-{\gamma(v)}({\bi r}-{\bi q})\times({\bi u}-{\bi v}),
\]
where the variables ${\bi r}$, ${\bi q}$, ${\bi u}$ and ${\bi v}$ are expressed in natural units. In general, the absolute value of both spins will be a function of the center of mass velocity ${\bi v}$, $S(v)$ and $S_{CM}(v)$,
with the values at rest  $S(0)=S_{CM}(0)=1/2$. The dynamics modifies the variables they depend and the final value will be determined during the dynamical process. 

From the electromagnetic point of view we need a natural unit for the electric charge. The usual assumption is to define the electric charge of the electron through the fine structure constant and this defines the permitivity of the vacuum $\epsilon_0$,
\[
\alpha=\frac{e^2}{4\pi\epsilon_0 c\hbar}=0.007297, \quad\rightarrow\quad e=1\,{\rm n.u.},\quad \frac{1}{4\pi\epsilon_0}=\alpha,
\]
The electron (and positron) has as intrinsic parameters $m=1$, $S=1/2$ and $e=\pm1$. 

What we have is to translate the international system of units to this natural system of units.
The relationship for the fundamental units of mass [M], length [L], time [T] and electric charge [Q] is:
\[
1\; {\rm n.u.} [M]=m_e=9.109534 \cdot10^{-31}{\rm Kg},\quad\rightarrow 1\;{\rm Kg}\equiv 1.09775\cdot10^{30}\;{\rm n.u.}
\]
\[
1\; {\rm n.u.} [L]=2R_0=\frac{\hbar}{mc}=3.86153\cdot10^{-13}{\rm m},\quad\rightarrow 1\; {\rm m}\equiv 2.58965\cdot10^{12}\;{\rm n.u.}
\]
\[
1\; {\rm n.u.} [T]=\tau_0=\frac{2R_0}{c}=6.44034\cdot10^{-22}\;{\rm s},\quad \rightarrow 1\; {\rm s}\equiv 7.76357\cdot10^{20}\;{\rm n.u.}
\]
\[
e=1 \;{\rm n.u.}\; [Q]=1.6021892\cdot10^{-19} {\rm C},\quad\rightarrow 1 \;{\rm C}\equiv 6.24146\cdot10^{18}\;{\rm n.u.}
\]
The electric field in the International System of units is expressed in V/m. According to the equivalence among units
\begin{equation}
1 \;{\rm V/m}=1\; {\rm m\, Kg\, s^{-2}\,C^{-1}},\quad 1\; {\rm V/m}=7.55676\cdot 10^{-19}\;{\rm n.u.} 
\label{voltM}
\end{equation}
The magnetic field is expressed in teslas. Since
\begin{equation}
1 \;{\rm T}=1\; {\rm Kg\, s^{-1}\,C^{-1}},\quad 1\; {\rm T}=2.26546\cdot 10^{-10}\;{\rm n.u.} 
\label{Teslanu}
\end{equation}

\subsection{Boundary Conditions}

To integrate the fourth-order system of differential equations of the point ${\bi r}(t)$ we need to supply the 12 values of ${\bi r}(0)$, ${\bi r}^{(1)}(0)$, ${\bi r}^{(2)}(0)$ and ${\bi r}^{(3)}(0)$, of the initial values of the point and their derivatives up to the third order. Nevertheless since the velocity ${\bi u}\equiv{\bi r}^{(1)}$ is of constant value $u=c$ and it is orthogonal to the acceleration  ${\bi u}\cdot{\bi a}={\bi r}^{(1)}\cdot{\bi r}^{(2)}=0$, we are left with only 10 independent boundary conditions.

To integrate the system of differential equations (\ref{eq:d2qdt2}) and (\ref{eq:d2rdt2}), we have to establish the appropriate boundary conditions for the positions and velocities of both points, the CC and the CM of each particle.
These 12 boundary values for the variables ${\bi r}(0)$, ${\bi u}(0)$, ${\bi q}(0)$ and ${\bi v}(0)$, are finally to be expressed in terms of 10 essential parameters because $|{\bi u}(0)|=1$ and ${\bi r}(0)-{\bi q}(0)$ is orthogonal to ${\bi u}(0)$. These 10 essential parameters are those parameters that define the relationship between the center of mass observer of the particle and any arbitrary inertial observer or laboratory observer who sees the CM of the particle moving at the speed ${\bi v}$. 

If $t^*$ and ${\bi r}^*$ are the time and position of the CC of the particle for the center of mass observer $O^*$ of this particle, and $t$ and ${\bi r}$ are the time and position of the CC for any arbitrary inertial observer, they are related by the Poincar\'e transformation:
\[
x^{\mu}=\Lambda^\mu_\nu x^{*\nu}+a^\mu,\quad x^\mu\equiv(ct,{\bi r}),\quad x^{*\mu}\equiv(ct^*,{\bi r}^*), \quad a^\mu\equiv(cb,{\bi d}),
\]
and $\Lambda=L({\bi v})R(\psi,\theta,\phi)$ is a general Lorentz transformation, as a composition of a rotation $R$, followed by a boost of velocity ${\bi v}$, or pure Lorentz transformation $L({\bi v})$.

We are going to describe now the rotation $R(\psi,\theta,\phi)$, the boost or pure Lorentz transformation $L({\bi v})$ with arbitrary velocity ${\bi v}$, the space translation  $T({\bi d})$ of displacement ${\bi d}$, and finally the time translation $T(b)$, of value $b$. 

\begin{figure}[!hbtp]\centering%
\includegraphics[width=7cm]{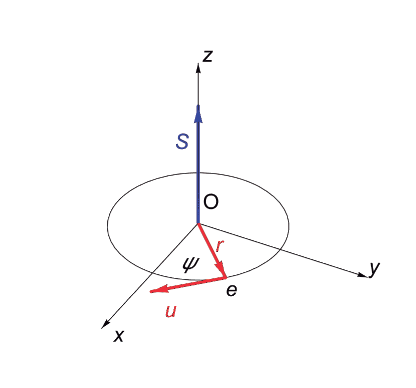}
\caption{Dirac particle in the center of mass reference frame, with the spin along the OZ axis. The initial position and velocity of the CC on the $XOY$ plane is determined by the phase  $\psi$. The radius of this motion is $R_0=1/2$, in natural units.} 
\label{fig2:initial}
\end{figure}

Let us first analyze the rotation. Let us assume we have the Dirac particle in the center of mass reference frame, as depicted in the figure {\bf\ref{fig2:initial}}, with the spin along $OZ$ axis.

If we initially have the CC at the position $(1/2,0,0)$ on the $OX$ axis, we rotate first an angle $\psi$, around the $OZ$ axis to modify the CC position. Next, we change the orientation of the spin by modifying the zenithal angle $\theta$, and azimuthal angle $\phi$, so that the arbitrary rotation is the composition of the rotations
\[
R(\psi,\theta,\phi)=R_{OZ}(\phi)R_{OY}(\theta)R_{OZ}(\psi),
\]
\[
R_{OZ}(\psi)=\pmatrix{\cos\psi&-\sin\psi&0\cr \sin\psi&\cos\psi&0\cr 0&0&1},
\]
\[
R_{OY}(\theta)=\pmatrix{\cos\theta&0&\sin\theta\cr 0&1&0\cr -\sin\theta&0&\cos\theta },\quad R_{OZ}(\phi)=\pmatrix{\cos\phi&-\sin\phi&0\cr \sin\phi&\cos\phi&0\cr 0&0&1}.
\]
If initially at time $t^*_0$, and in natural units the CC is at the point $(1/2,0,0)$, then the position, velocity and acceleration of the CC, before the rotation, since $c=1$ and $a=c^2/R_0=2$, are:
\[
{\bi r}^*(t^*_0)=\pmatrix{1/2\cr 0\cr 0},\quad {\bi u}^*(t^*_0)=\pmatrix{0\cr -1\cr 0}, \quad {\bi a}^*(t^*_0)=\pmatrix{-2\cr 0\cr 0}.
\]
These variables, after the rotation, become ${\bi r}^*_0=R_{OZ}(\phi)R_{OY}(\theta)R_{OZ}(\psi){\bi r}^*(t^*_0)$, and the same for ${\bi u}^*(t^*_0)$ and ${\bi a}^*(t^*_0)$, at the same time $t^*_0$, and thus:
 \begin{equation}
{\bi r}^*_0=\frac{1}{2}\pmatrix{\cos\theta\cos\phi\cos\psi-\sin\phi\sin\psi\cr \cos\theta\sin\phi\cos\psi+\cos\phi\sin\psi\cr
-\sin\theta\cos\psi},
 \label{eq:valoresinicialesr}
 \end{equation}
  \begin{equation}
  {\bi u}^*_0=\pmatrix{\cos\theta\cos\phi\sin\psi+\sin\phi\cos\psi\cr \cos\theta\sin\phi\sin\psi-\cos\phi\cos\psi\cr
-\sin\theta\sin\psi},
 \label{eq:valoresinicialesu}
 \end{equation}
\begin{equation}
{\bi a}^*_0=-2\pmatrix{\cos\theta\cos\phi\cos\psi-\sin\phi\sin\psi\cr \cos\theta\sin\phi\cos\psi+\cos\phi\sin\psi\cr
-\sin\theta\cos\psi}=-4{\bi r}^*_0.
 \label{eq:valoresinicialesac}
\end{equation}
The Lorentz transformation $L({\bi v})$ is given by the $4\times4$ matrix
 \begin{equation}
L({\bi v})=\pmatrix{\gamma&\gamma{v_x}& \gamma{v_y}& 
\gamma{v_z}\cr 
\gamma{v_x}&1+{\displaystyle v_x^2\gamma^2\over\displaystyle\gamma+1}&{\displaystyle v_xv_y\gamma^2\over\displaystyle\gamma+1}&{\displaystyle v_xv_z\gamma^2\over\displaystyle\gamma+1}\cr 
\gamma{v_y}&{\displaystyle v_yv_x\gamma^2\over\displaystyle\gamma+1}& 1+{\displaystyle v_y^2\gamma^2\over\displaystyle 
\gamma+1}&{\displaystyle v_yv_z\gamma^2\over\displaystyle\gamma+1}\cr 
\gamma{v_z}&{\displaystyle v_zv_x\gamma^2\over\displaystyle \gamma+1}&{\displaystyle v_zv_y\gamma^2\over\displaystyle \gamma+1}&1+{\displaystyle v_z^2\gamma^2\over\displaystyle \gamma+1}\cr},
 \label{eq:Tdev}
 \end{equation} 
where $\gamma\equiv(1-v_x^2-v_y^2-v_z^2)^{-1/2}$. Finally the two translations $T({\bi d})$ and $T(b)$.
The corresponding time $t_0$ and position ${\bi r}_0$ of the CC of the same event, for any arbitrary inertial observer in natural units are:
\begin{equation}
t_0=\gamma\left(t^*_0+{{\bi v}\cdot{\bi r}^*_0}\right)+b,\quad {\bi r}_0={\bi r}^*_0+\gamma{\bi v}t^*_0+\frac{\gamma^2}{1+\gamma}{({\bi v}\cdot{\bi r}^*_0){\bi v}}+{\bi d},
\label{eq:transt0r0}
\end{equation}
where $b$ is the time translation, ${\bi d}$ is the space translation in natural units and ${\bi v}$ is the velocity of the center of mass observer $O^*$ as measured by the observer $O$. It represents, therefore, the velocity of the center of mass of the particle for the arbitrary laboratory observer $O$.

If the initial instant to integrate the equations in the reference frame of the Laboratory $O$, is the time $t_0=0$, this corresponds to
$\gamma t^*_0=-\gamma {\bi v}\cdot{\bi r}^*_0-b$, for the center of mass observer of the particle, and therefore the initial position ${\bi r}_0$ of the CC for the laboratory observer at the initial instant  $t_0=0$, instead of (\ref{eq:transt0r0}) is:
\begin{equation}
{\bi r}_0={\bi r}^*_0-\frac{\gamma}{1+\gamma}{({\bi v}\cdot{\bi r}^*_0){\bi v}}-b{\bi v}+{\bi d}.
\label{ecuac:r0}
\end{equation}
For the velocity, the transformation is: 
 \begin{equation}
{\bi u}_0=\frac{{\bi u}^*_0+\gamma{\bi v}+\frac{\gamma^2}{(1+\gamma)}{({\bi v}\cdot{\bi u}^*_0){\bi v}}}{\gamma(1+{\bi v}\cdot{\bi u}^*_0)},
 \label{ecuac:u0}
 \end{equation} 
and for the acceleration the transformation in natural units is:
 \begin{equation}
{\bi a}_0=\frac{(1+{\bi v}\cdot{\bi u}^*_0){\bi a}^*_0-({\bi v}\cdot{\bi a}^*_0){\bi u}^*_0-\frac{\gamma}{(1+\gamma)}({\bi v}\cdot{\bi a}^*_0){\bi v}}{\gamma^2(1+{\bi v}\cdot{\bi u}^*_0)^3},
 \label{ecuac:a0}
 \end{equation}
where ${\bi r}^*_0$, ${\bi u}^*_0$ and ${\bi a}^*_0$ are given above in (\ref{eq:valoresinicialesr}), (\ref{eq:valoresinicialesu}) and (\ref{eq:valoresinicialesac}), respectively.

\begin{figure}[!hbtp]\centering%
\includegraphics[width=7cm]{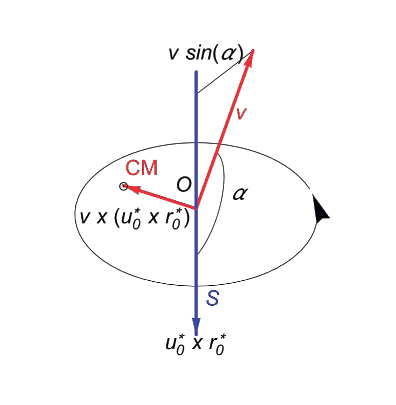}
\caption{Initial position of the CM when the velocity of the CM
${\bi v}$ and the spin direction form an angle $\alpha$. It is perpendicular to the vectors ${\bi v}$ and ${\bi u}^*_0\times{\bi r}^*_0$, of distance to the center $O-CM=(v/2)\sin\alpha$, in natural units, and is independent of the initial position and velocity of the CC. In general, the separation between both centers CC and CM, is in the interval $|{\bi r}-{\bi q}|\in[1/2-(v/2)\sin\alpha,1/2+(v/2)\sin\alpha]$, in natural units, so that this separation is not a constant of the motion.} 
\label{fig3:posCMini}
\end{figure}
With these values we have the initial boundary conditions for the variables ${\bi r}_0$, ${\bi u}_0$ and ${\bi v}_0={\bi v}$. What we have left is ${\bi q}_0$, the initial position of the CM in this reference frame.

The definition of the center of mass position of the Dirac particle in any reference frame at any time is given in (\ref{defposCM}) and in natural units
\[
{\bi q}(t)={\bi r}(t)+\frac{1-{\bi v}(t)\cdot{\bi u}(t)}{{\bi a}(t)^2}\,{\bi a}(t),
\]
then the initial value of the CM ${\bi q}_0$, in the laboratory frame is:
\begin{equation}
{\bi q}_0={\bi r}_0+\frac{1-{\bi v}\cdot{\bi u}_0}{{{\bi a}_0}^2}\,{\bi a}_0,
\label{eq:iniPOSCM}
\end{equation}
in terms of the position, velocity and acceleration of the CC in this reference frame and of the velocity of the CM, all these variables defined at the initial time $t_0=0$.

Since ${\bi a}^*_0=-4{\bi r}^*_0$, taking the squared of (\ref{ecuac:a0}) we get
\[
{{\bi a}_0}^2=\frac{4}{\gamma^4(1+{\bi v}\cdot{\bi u}^*_0)^4},
\]
From (\ref{ecuac:u0}) the term
\[
1-{\bi v}\cdot{\bi u}_0=\frac{1}{\gamma^2(1+{\bi v}\cdot{\bi u}^*_0)}\;\Rightarrow\;
\frac{1-{\bi v}\cdot{\bi u}_0}{{\bi a}_0^2}{\bi a}_0=\frac{1}{4}{\gamma^2}(1+{\bi v}\cdot{\bi u}^*_0)^3{\bi a}_0,
\]
and if we substitute for ${\bi a}_0$ the expression (\ref{ecuac:a0}), where ${\bi a}^*_0=-4{\bi r}^*_0$, if we use (\ref{ecuac:r0}) the initial position (\ref{eq:iniPOSCM}) of the CM, ${\bi q}_0$, is
\begin{equation}
{\bi q}_0
={\bi v}\times({\bi u}^*_0\times{\bi r}^*_0)-b{\bi v}+{\bi d}.
\label{inicialcondCM}
\end{equation}

As a summary, the boundary conditions at $t_0=0$, in the laboratory frame, are:
 \begin{equation}
{\bi r}_0={\bi r}^*_0-\frac{\gamma}{1+\gamma}{({\bi v}\cdot{\bi r}^*_0){\bi v}}-b{\bi v}+{\bi d}, 
 \label{eq:initialcond1}
 \end{equation}
 \begin{equation}
 {\bi u}_0=\frac{{\bi u}^*_0+\gamma{\bi v}+\frac{\gamma^2}{(1+\gamma)}{({\bi v}\cdot{\bi u}^*_0){\bi v}}}{\gamma(1+{\bi v}\cdot{\bi u}^*_0)},
 \label{eq:initialcond3}
 \end{equation}
 \begin{equation}
{\bi q}_0={\bi v}\times({\bi u}^*_0\times{\bi r}^*_0)-b{\bi v}+{\bi d},
 \label{eq:initialcond2}
 \end{equation}
 \begin{equation} 
 {\bi v}_0={\bi v}.
  \label{eq:initialcond4}
 \end{equation}
These boundary conditions are expressed in terms of 10 essential parameters: the space and time parameters ${\bi d}$ and $b$ that define the initial location of the CC and CM and the initial instant of the integration, respectively. The three parameters ${\bi v}$ that define the initial velocity of the CM and finally the three angles $\psi$, the initial phase of the CC, and $\theta$ and $\phi$ that define the initial orientation of the spin in the center of mass frame.

In general, the initial position of the CM for the laboratory observer, given in (\ref{eq:initialcond2}), is contained in the zitterbewegung plane (with ${\bi d}=0=b$),
is independent of the initial position of the CC, ${\bi r}^*_0$,
and is depicted in the figure {\bf\ref{fig3:posCMini}}.
The distance to the center is $(v/2)\sin\alpha$.
If ${\bi v}$ is perpendicular to the ziterbewegung plane, ${\bi v}\cdot{\bi r}^*_0={\bi v}\cdot{\bi u}^*_0=0$, and the CM is at the center of the circle and becomes equidistant of the trajectory of the CC. If ${\bi v}$, has a different orientation, because $R_0=1/2$, the separation between both centers CC and CM, $|{\bi r}-{\bi q}|\in[1/2-(v/2)\sin\alpha,1/2+(v/2)\sin\alpha]$, in natural units, and is not a constant of the motion. 

We must remark that the CC ${\bi r}_0$ is the Poincar\'e transformed of the point ${\bi r}^*_0$, but 
${\bi q}_0$ is not the Poincar\'e transformed of the CM ${\bi q}^*_0$ in $O^*$. It is the CM of the particle in the laboratory frame $O$, at the same time $t_0$ than ${\bi r}_0$, while ${\bi r}^*_0$ and ${\bi q}^*_0$ are considered simultaneous in the frame $O^*$. Simultaneous events in one frame are not simultaneous in another. The point ${\bi q}_0$ represents the location of the CM of the particle simultaneous with ${\bi r}_0$, at the laboratory observer.

\subsection{Dirac particle in a uniform magnetic field}
\label{sec:uniformB}
We are going to describe the motion of the Dirac particle in a uniform magnetic field. To draw
in the same figure the cyclotron motion of the center of mass, that has a radius of order of $10^{-6}$m in a magnetic field of 5 Teslas, and the zitterbewegung motion of the CC around the CM of much smaller radius $R_0\simeq10^{-13}$m, we have to consider a high velocity electron and a very large magnetic field, to appreciate both motions in this figure.

In a uniform magnetic field along $OZ$ axis ${\bi B}=B{\bi e}_z$, the dynamical equation (\ref{eq:d2qdt2}) in natural units becomes
\[
\frac{dv_x}{dt}=\frac{KB}{\gamma(v)}\left[u_y-v_x(v_xu_y-v_yu_x)\right],
\]
\[
\frac{dv_y}{dt}=\frac{KB}{\gamma(v)}\left[-u_x-v_y(v_xu_y-v_yu_x)\right],
\]
\[
\frac{dv_z}{dt}=\frac{KB}{\gamma(v)}\left[-v_z(v_xu_y-v_yu_x)\right],
\]
\begin{equation}
 K=\frac{e\hbar}{2m^2c ^2}=1.13273\cdot10^{-10}\; {\rm T}^{-1}, 
 \label{constantK}
\end{equation}
where  the magnetic field $B$ is given in Teslas.

\begin{figure}[!hbtp]\centering%
 \includegraphics[width=.5\textwidth]{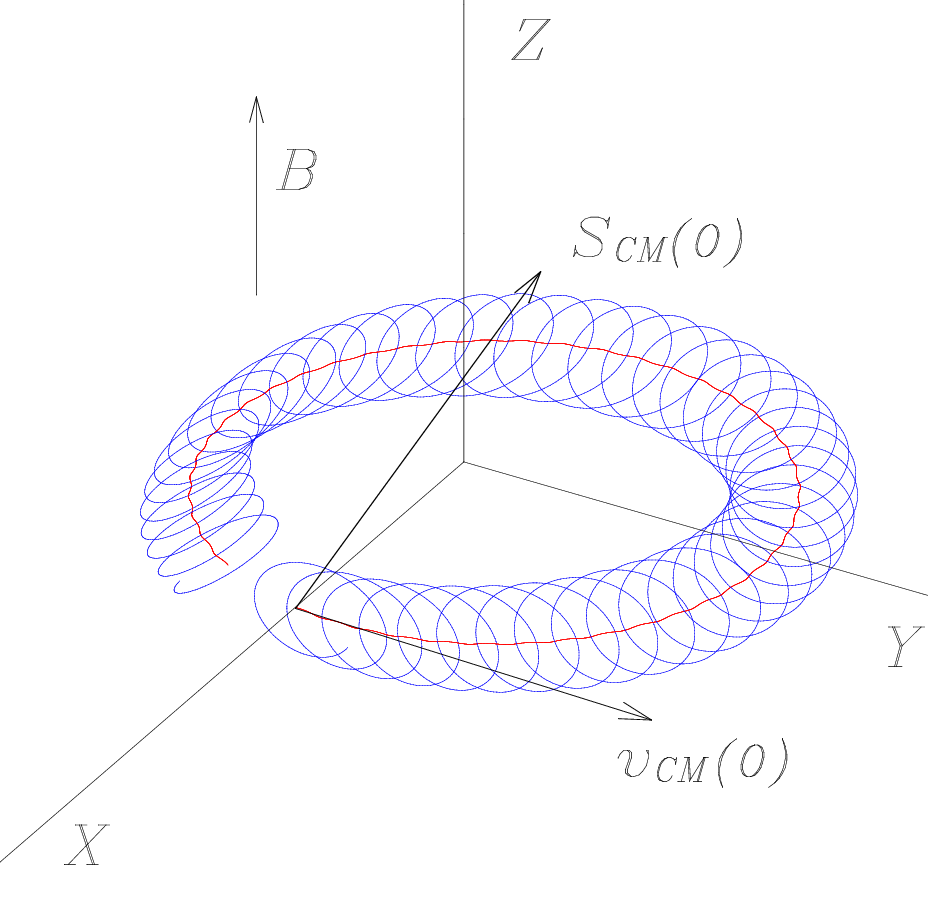}
 \caption{\small{Cyclotron motion of the CM (red) and zitterbewegung motion of the CC (blue), with $v/c=0.1$, spin orientation $\theta=30^\circ$, $\phi=90^\circ$ and
$B=2\cdot10^8$.}}
\label{fig:cyclo01}
\end{figure}

For the initial conditions in natural units
\[
\theta=30^\circ,\phi=90^\circ,\quad v_y(0)=0.1,\quad d_x=5,\quad b=0,
\]
and a magnetic field of $B=2\cdot10^8$, we get the cyclotron motion of the CM of radius $R_c\simeq5$ and zitterbewegung motion of the CC of internal radius $1/2$, of the figure {\bf\ref{fig:cyclo01}}.

\begin{figure}[!hbtp]\centering%
\hspace{1cm}\includegraphics[width=.40\textwidth]{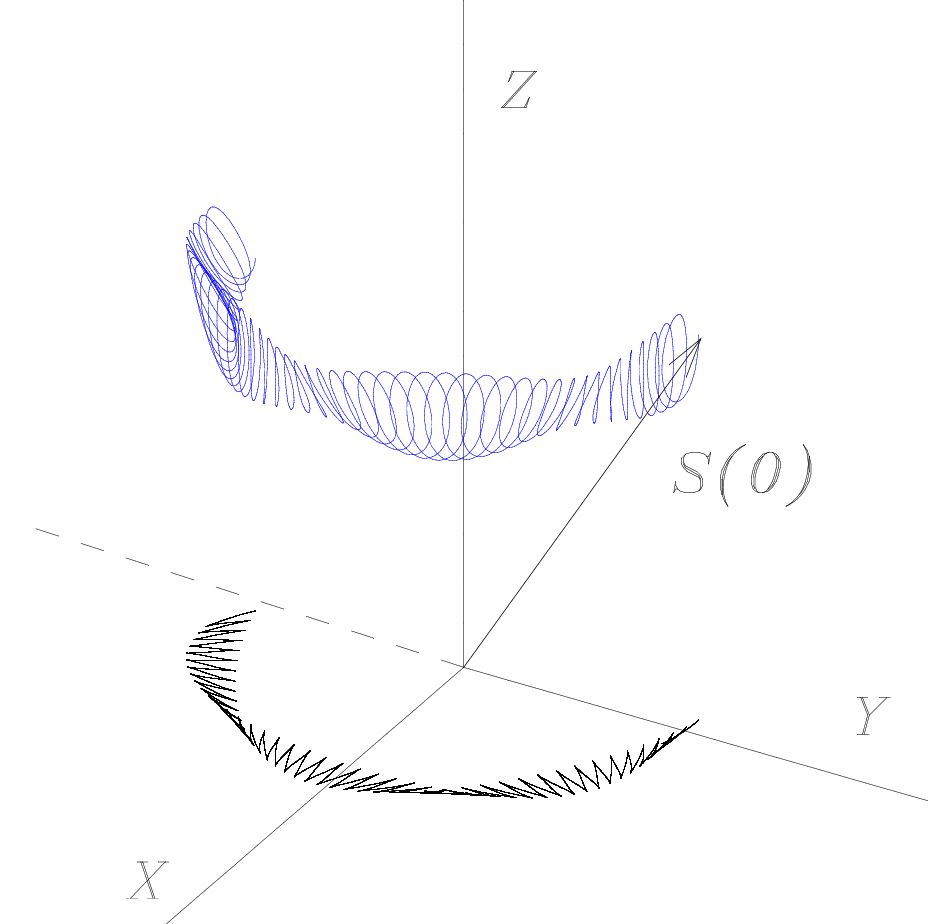}\hspace{0.5cm}\includegraphics[width=.40\textwidth]{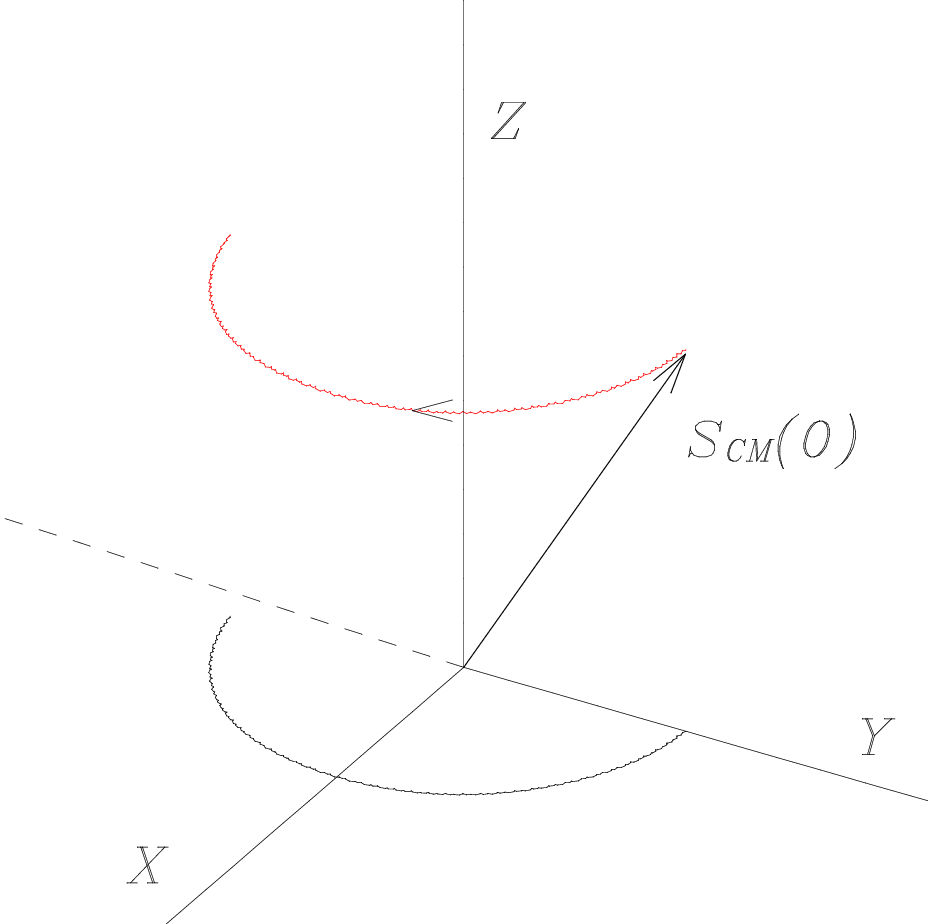}\caption{Evolution of the end point of the CC spin ${\bi S}$ (blue) and its projection onto the $XOY$ plane(black) and the evolution of the CM spin ${\bi S}_{CM}$ (red) and its projection onto the $XOY$ plane(black), corresponding to the same time of the evolution of figure {\ref{fig:cyclo01}}. Both spins precess backwards with the angular velocity $-\omega_c/2$.}
\label{fig:spindynamics}
\end{figure}
 
We also draw in figure {\bf\ref{fig:spindynamics}}, the evolution of both spins, the CC spin ${\bi S}$ and the CM spin ${\bi S}_{CM}$ during the same time of the evolution of the figure {\bf\ref{fig:cyclo01}}. The CC spin moves in the orthogonal direction to the linear momentum, because it satisfies Dirac's dynamical equation $d{\bi S}/dt={\bi p}\times{\bi u}$. Here the time derivative of the spin is always orthogonal to the direction of the linear momentum and this produces the spin precession. We see that the CM spin behaves like the average of the CC spin. Both spins precesses backwards with the angular velocity $\omega_s$, half the cyclotron angular velocity $\omega_s=-\omega_c/2$. This is in contradiction with the assumption of the experimental works \cite{Dehmelt1981}-\cite{Hanneke2008}, where it is assumed that the spin precession angular velocity is $\omega_s=g\omega_c/2$.

\begin{figure}[!hbtp]\centering%
 \hspace{1cm}\includegraphics[width=.40\textwidth]{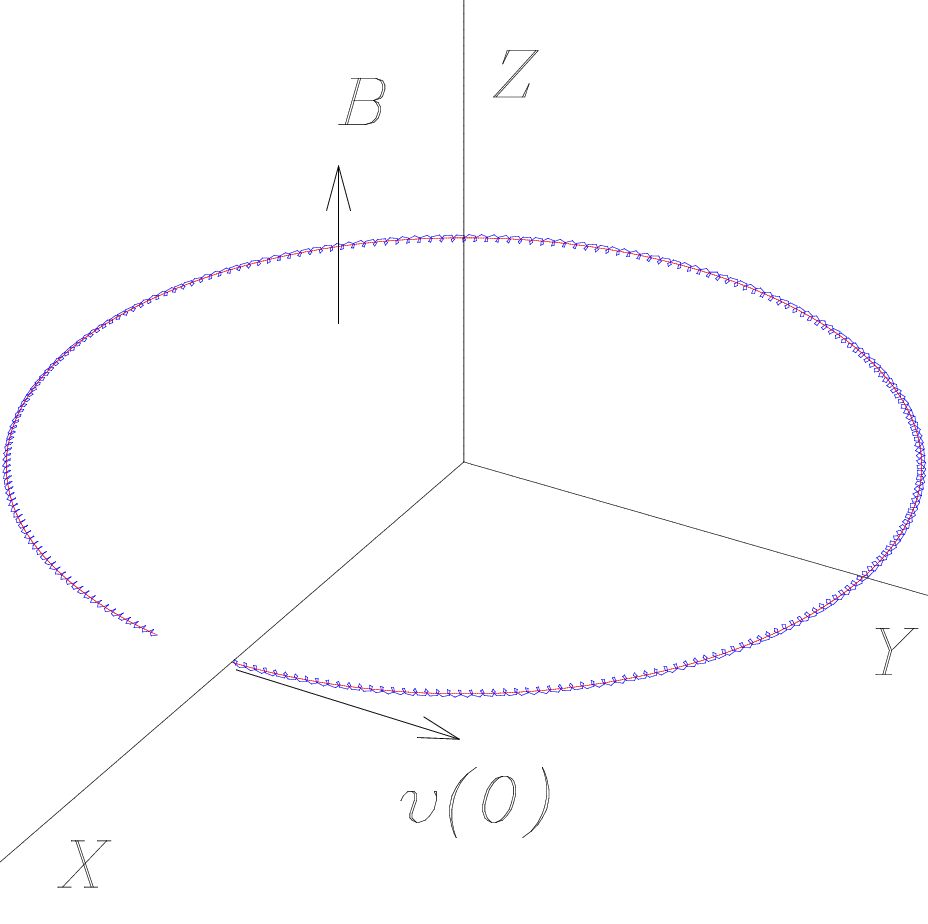}\hspace{0.5cm}\includegraphics[width=.40\textwidth]{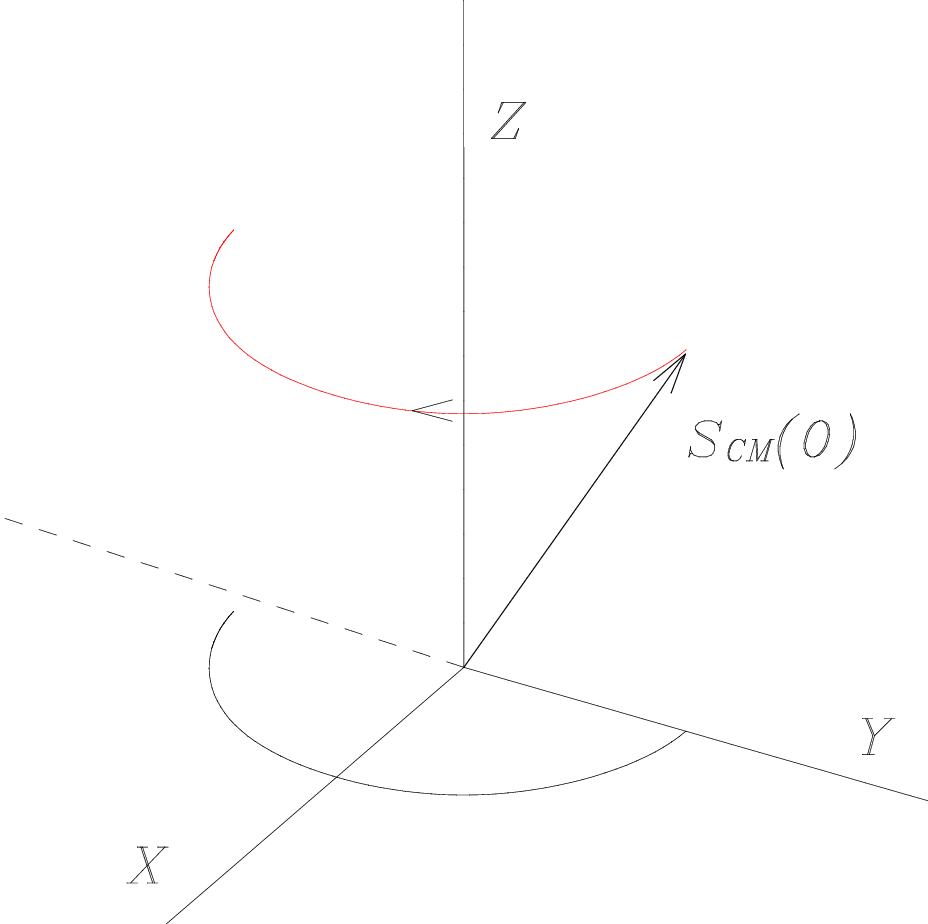}\caption{Cyclotron motion of radius $R_c\approx100$ of the CC and CM, and the evolution of the CM spin ${\bi S}_{CM}$ (red) with the same orientation as in the figure \ref{fig:spindynamics}, and its projection onto the $XOY$ plane(black), during the same time as the cyclotron motion on the left. The initial velocity is $v=0.1$ and the magnetic field $B=9\cdot10^6$. As before, the CM spin ${\bi S}_{CM}$ precesses backwards with Larmor's angular velocity $-\omega_c/2$.}
\label{fig:spin100}
\end{figure}

In the figure {\bf\ref{fig:spin100}} we draw the same cyclotron motion, with the same spin orientation, $\theta=30^\circ$, $\phi=90^\circ$, but now the magnetic field is smaller than in the previous integration $B=9\cdot10^6$. The initial velocity $v/c=0.1$ is the same, but produces a bigger cyclotron radius $R_c\approx100$, in dimensionless units. This scale corresponds to a cyclotron motion of radius $R_c\approx 10^{-11}$m. The trajectories of the CC and CM with a separation of order $R_0\approx10^{-13}$m, appear too close to be able to distinguish them separately. Nevertheless the spin evolution is exactly the same as before. It precesses backwards with the angular velocity $\omega_s=-\omega_c/2$.

With a magnetic field of 5 Teslas, the inicial velocity like in the other examples $v/c=0.1$, the cyclotron radius is $R_c\approx 10^8$ in natural units, but the motion of the spin is the same as in the above figure {\bf\ref{fig:spin100}}, where the ${\bi S}_{CM}$, precesses backwards with the angular velocity $-\omega_c/2$, because Dirac's spin ${\bi S}$ satisfies the dynamical equation
$d{\bi S}/dt={\bi p}\times{\bi u}$. 

\subsection{Definition of magnetic moment for the Dirac particle}
\label{secDefmu}

According to Jackson \cite{Jackson}, the magnetic moment of a current density ${\bi j}(t,{\bi r})$ with respect to a point, which we take as the origin of our reference frame, is defined as the vector:
\begin{equation}
\bmu=\frac{1}{2}\int {\bi r}\times {\bi j}(t,{\bi r})dV.
\label{eq:Jackson}
\end{equation}
where $dV$ is the volume element around the point ${\bi r}$. In the case of a flat closed circuit of area $A$ and intensity $I$, the above definition leads for the absolute value of $\mu$, to $\mu=IA$.

For a point particle of charge $e$, located at the point ${\bi r}$, and moving at the velocity 
${\bi v}$, the instantaneous magnetic moment with respect to the origin of the reference frame,  compatible with the above definition (\ref{eq:Jackson}), is:
\begin{equation}
\bmu=\frac{1}{2}e{\bi r}\times{\bi v},
\label{defMuJack}
\end{equation}
since $e{\bi v}$, plays the role of ${\bi j}(t,{\bi r})dV$. If the point particle has mass $m$, the relativistic angular momentum with respect to the origin of the reference frame is
\begin{equation}
{\bi J}={\bi r}\times{\bi p}=m\gamma(v){\bi r}\times{\bi v},\quad\Longrightarrow\quad \bmu=\frac{e}{2m\gamma(v)}{\bi J},
\label{munonrel}
\end{equation}
and we find a relationship between the instantaneous magnetic moment and the angular momentum with respect to the same point, for the spinless point particle.

For compound mechanical systems like nucleons or atomic nuclei the relationship between the total magnetic moment $\bmu$ and its total angular momentum ${\bi J}$ with respect to the center of mass takes the form 
\[
\bmu=g\frac{e}{2m}{\bi J},
\]
where $e$ represents the total charge and $m$ the total mass of the system. The dimensionless magnitude $g$ is called the {\it gyromagnetic ratio} of the system.

In the Dirac particle model of charge $e$, of figure {\bf{4}}, moving in circles of radius $R_0$ at the speed $c$, the period of this motion is $T_0={2\pi R_0}/{c}$, and this motion is equivalent to a current of intensity $I=e/T_0$. The area enclosed is $A=\pi R_0^2$, and the absolute value of the magnetic moment, according to the definition (\ref{eq:Jackson}), is:
\[
\mu_B=IA=\frac{e\hbar}{2m},
\]
which is called Bohr's magneton. When expressed this $\mu_B$ in terms of the spin with respect to the center of mass $S=\hbar/2$, it takes the form:
\[
\mu_B=2\frac{e}{2m}\frac{\hbar}{2}=g\frac{e}{2m}S,\quad\Rightarrow\quad g=2,
\]
and the quantity $g=2$, which is absent in the definition of the magnetic moment for the spinless electron (\ref{munonrel}), is called the gyromagnetic ratio of the electron. This expression for the magnetic moment and its relation with the angular momentum for the relativistic Dirac particle is different than in the case of the relativistic spinless point particle (\ref{munonrel}).

That $g\neq2$ for compound systems reflects the fact that these objects are not elementary spinning particles and this difference shows the existence of an internal structure.

Therefore, for the Dirac elementary particle, the relationship between its magnetic moment with respect to the CM and the spin with respect to the CM, should include the factor $g=2$, and, instead of (\ref{munonrel}), will be
\begin{equation} 
 \bmu_{CM}=\frac{e}{m\gamma(v)}{\bi S}_{CM}.
\label{eq:defMu}
\end{equation}  
We take, for the Dirac particle, as the exact relationship between the instantaneous magnetic moment and the spin with respect to the center of mass, the above definition (\ref{eq:defMu})
as predicted by Dirac equation.

Using the expression of ${\bi S}_{CM}$ in (\ref{eq:spinCM}) the appropriate definition of the instantaneous magnetic moment with respect to the center of mass of the Dirac particle, in terms of the classical variables which describe its motion, will be:
\begin{equation}
{\bmu}_{CM}=-e({\bi r}-{\bi q})\times({\bi u}-{\bi v}),
\label{eq:mupointCM}
\end{equation}
which lacks the factor $1/2$ of the corresponding definition for the spinless point particle (\ref{defMuJack}). 
In the case of the Dirac particle at rest, the instantaneous magnetic moment and its average value are the same, and the absolute value is,
\[
\mu=eR_0c=\frac{e\hbar}{2m}=\mu_B.
\]
In natural units for the electron $\mu_B=1/2$.
\subsection{Definition of electric dipole moment for the Dirac particle}
\label{secDefED}
The definition of the instantaneous electric dipole moment for a system of point particles of charges $q_i$, located at the respective points ${\bi r}_i$, with respect to the origin of the reference frame, is
\[
{\bi d}=\sum_i q_i{\bi r}_i.
\]
In the case of the Dirac particle, the electric dipole moment with respect to the CM is:
\begin{equation}
{\bi d}_{CM}=e({\bi r}-{\bi q}).
\label{elecdipolMom}
\end{equation}
For the Dirac particle at rest, this dipole is not a constant of the motion. Its average value during a turn of the charge is 0, but the absolute value is the constant $d_{CM}=eR_0$. Therefore, the measurement of the electric dipole moment with respect to the center of mass is equivalent to the measurement of the average separation between the CM and CC, and the existence of this dipole is not related to any asymmetry of the possible charge distribution. In chapter 6 of \cite{Rivasbook} it is shown that the above classical definition of electric dipole moment (\ref{elecdipolMom}) leads, when quantizing this model, to Dirac's electric dipole moment operator. We arrive to this result by making use of Hestenes {\it geometric algebra} \cite{Hestenes}. It is also shown in that chapter that the absolute value of this dipole gives rise to Darwin's term of Dirac's Hamiltonian. In natural units $d_{CM}=1/2$.

\section{Dirac particle in a Penning trap}
\label{DiracPenning}

In the Penning trap of Section {\bf\ref{sec:pening}}, the electromagnetic field is
\[
{\bi E}=\frac{V_0R_0}{2d^2}(r_x,r_y,-2r_z),\quad {\bi B}=(0,0,B),
\]
with $V_0$ in Volts, $B$ in Teslas and the magnetron dimension $d$ in meters. 
The fields are defined at the CC position ${\bi r}$, of the electron, in natural units.

We are going to integrate the above dynamical equations with the data of the experiment performed by Gabrielse and Dehlmet \cite{Dehmelt1981}, where the fields and parameters take the values:
\[
V_0=10.22\, {\rm V},\quad B=5.872\, {\rm T}, \quad d=0.00335\, {\rm m}.
\]

If we express the variables in the fields in terms of dimensionless variables in natural units, the dynamical equations of the Dirac particle in the Penning trap are:
\[
\frac{dq_x}{dt}=v_x,\quad\frac{dq_y}{dt}=v_y,\quad
\frac{dq_z}{dt}=v_z,
\]
\[
\frac{dv_x}{dt}=\frac{1}{\gamma(v)}\left[C(r_x-v_x(v_xr_x+v_yr_y-2v_zr_z))+K(u_y+v_xu_y-v_yu_x)\right],
\]
\[
\frac{dv_y}{dt}=\frac{1}{\gamma(v)}\left[C(r_y-v_y(v_xr_x+v_yr_y-2v_zr_z))+K(-u_x+v_xu_y-v_yu_x)\right],
\]
\[
\frac{dv_z}{dt}=\frac{1}{\gamma(v)}\left[-C(2r_z+v_z(v_xr_x+v_yr_y-2v_zr_z))+K(v_xu_y-v_yu_x)\right],
\]
\[
\frac{dr_x}{dt}=u_x,\quad\frac{dr_y}{dt}=u_y,\quad\frac{dr_z}{dt}=u_z,
\]
\[
\frac{du_x}{dt}=\frac{1-v_xu_x-v_yu_y-v_zu_z}{(q_x-r_x)^2+(q_y-r_y)^2+(q_z-r_z)^2}(q_x-r_x),
\]
\[
\frac{du_y}{dt}=\frac{1-v_xu_x-v_yu_y-v_zu_z}{(q_x-r_x)^2+(q_y-r_y)^2+(q_z-r_z)^2}(q_y-r_y),
\]
\[
\frac{du_z}{dt}=\frac{1-v_xu_x-v_yu_y-v_zu_z}{(q_x-r_x)^2+(q_y-r_y)^2+(q_z-r_z)^2}(q_z-r_z),
\]
\noindent
where the dimensionless coefficients are:
\begin{equation}
C=\frac{e\hbar V_0}{4d^2m^2c^3}=-1.72043\cdot 10^{-13},
\label{coeffC}
\end{equation}
\begin{equation}
K=\frac{e\hbar B}{2m^2c^2}=-6.65139\cdot 10^{-10}.
\label{coeffK}
\end{equation}
and
\[
\gamma(v)=(1-v_x^2-v_y^2-v_z^2)^{-1/2}. 
\]

The predicted features of the classical description, where the internal motion of the CC is a very high frequency motion of $\omega_0\simeq 10^{21}$ s$^{-1}$, explain that when we try to make a measurement, for example, the instantaneous value of any magnitude such as positions and velocities of points they are impossible to determine. Each measurement takes time, so we understand that the classical measurement of any observable is the temporal average value of a sufficiently large number of local measurements.

In this integration what we are going to measure first is the absolute value of the electric dipole moment (\ref{elecdipolMom}), which is going to be compared with its theoretical value $1/2$, in natural units. 
This separation vector between the CC and CM is not a constant vector. The time average value is $1/2$ for the free particle, and will be greater than $1/2$ when moving. To determine its value, we shall calculate the time average value of this expression during a complete number of turns of the CC. With the electron at rest, the time to complete a turn is $T_0=2\pi R_0/c=\pi$, in natural units. If the CM of the Dirac particle is moving at the speed $v$, then the time during a complete turn is $T=\gamma(v)T_0=\gamma(v)\pi$. Therefore, the time average value of this separation for a Dirac particle $|{\bi r}-{\bi q}|$, after $n$ turns of the CC, is defined as
\[
d_{CM}(v)=\frac{1}{\pi n\gamma(v)}\int_{0}^{\pi n\gamma(v)}\,|{\bi r}-{\bi q}|\,dt.
\]
 
The other measurement is the absolute value of the magnetic moment with respect to the center of mass. In natural units is
\[
\mu_{CM}=|({\bi r}-{\bi q})\times({\bi u}-{\bi v})|.
\]
This measurement will be for the free particle $1/2$ and it is going to be $\le1/2$ for interacting particles, and not greater than $1/2$, as is suggested by the aforementioned magnetic moment experiments. 

We compute these magnitudes for $n=100$ turns. For the velocity of the center of mass of the electron we take the three values $v_y=0.1,\; 0.01,\;0.001$, which correspond to a cyclotron radius of value $2.9\cdot10^{-5},\;2.9\cdot10^{-6},\;2.9\cdot10^{-7}\;$m, respectively. For every velocity we take seven spin orientations $\alpha$, with respect to the velocity vector, of values
$\alpha=90^\circ,\; 80^\circ,\;60^\circ,\;45^\circ,\;30^\circ,\;10^\circ,\;0^\circ\;$. We reproduce the following three tables corresponding to the different velocities.

The table for $v=0.1$, \\
  \begin{tabular}{|c|c|l|c|c|}
\hline 
$\alpha$ & $v$ & $(v/2)\sin\alpha$ & $d_{CM}$ & $\mu_{CM}$ \\ 
\hline 
$90^\circ$ & 0.1 & 0.05 &0.50249713362 &0.494996859275  \\
\hline 
$80^\circ$ & 0.1 & 0.049240385 & 0.50242142065 &0.495073021263 \\ 
\hline 
$60^\circ$ & 0.1 & 0.04330127 & 0.50187181145 & 0.495622246610 \\ 
\hline 
$45^\circ$ & 0.1 & 0.035355335 & 0.50124681475 & 0.496246062917 \\ 
\hline 
$30^\circ$ & 0.1 & 0.025 & 0.50062182705 & 0.496869096535 \\ 
\hline 
$10^\circ$ & 0.1 & 0.0086824 & 0.50007221173 & 0.497416338231 \\ 
\hline 
$0^\circ$ & 0.1 & 0 & 0.49999682765 &  0.497491349886\\ 
\hline  
\end{tabular} 

\vspace{0.5cm}

The table for  $v=0.01$, \\
\begin{tabular}{|c|c|l|c|c|}
\hline
$\alpha$ & $v$ & $(v/2)\sin\alpha$ & $d_{CM}$ & $\mu_{CM}$ \\ 
\hline 
$90^\circ$ & 0.01 & 0.005 &0.50002499873& 0.4999499999486\\
\hline 
$80^\circ$ & 0.01 & 0.0049240385 & 0.50002424525&0.4999507533125 \\ 
\hline 
$60^\circ$ & 0.01 & 0.004330127 & 0.50001874445 & 0.499956249432 \\
\hline 
$45^\circ$ & 0.01 & 0.003535533 &0.50001249915 & 0.499962499314 \\
\hline
$30^\circ$ & 0.01 & 0.0025 & 0.50000624932 & 0.499968749119 \\ 
\hline 
$10^\circ$ & 0.01 & 0.00086824 & 0.50000075325 & 0.499974245045 \\ 
\hline 
$0^\circ$ & 0.01 & 0 &  0.49999999945 & 0.499974998845 \\ 
\hline 
\end{tabular} 

\vspace{0.5cm}

Finally the table for $v=0.001$,\\
\begin{tabular}{|c|c|l|c|c|}
\hline 
$\alpha$ & $v$ & $(v/2)\sin\alpha$ & $d_{CM}$ & $\mu_{CM}$ \\ 
\hline 
$90^\circ$ & 0.001 & 0.0005 & 0.50000024953& 0.499999499964 \\
\hline 
$80^\circ$ & 0.001 & 0.00049240385 &0.50000023845& 0.499999507486 \\
\hline 
$60^\circ$ & 0.001 & 0.0004330127 &0.50000018385 & 0.499999562447 \\ 
\hline 
$45^\circ$ & 0.001 & 0.0003535533 &0.50000012157 & 0.499999624947 \\ 
\hline 
$30^\circ$ & 0.001 & 0.00025 & 0.50000006225& 0.499999687447 \\ 
\hline 
$10^\circ$ & 0.001 & 0.000086824 & 0.50000000742 & 0.499999742408 \\ 
\hline
$0^\circ$ & 0.001 & 0 & 0.49999999995 & 0.499999749947 \\ 
\hline 
\end{tabular} 
 
 \vspace{0.5cm}
The expected measurement of the exact electric dipole moment in natural units is $1/2$. Nevertheless, this time average value is not a constant of the motion and depends on the value $v_\alpha=(v/2)\sin\alpha$, which is related to the initial separation between the CM and CC, as we mentioned before (see figure {\bf\ref{fig3:posCMini}}). The smaller this magnitude $v_\alpha$, smaller is the time average value of the absolute value of the dipole $d_{CM}$, which is above and approaching to the theoretical value of $1/2$.

For the time-average value of the magnetic moment with respect to the CM, we also see that the absolute value is not constant, it also depends on the value $v_\alpha$, and when the velocity is smaller it approaches the theoretical value $1/2$ of the free particle. The magnetic moment decreases as the CM velocity increases. The absolute value of the magnetic moment with respect to the center of mass is always slightly smaller than the theoretical value predicted by Dirac.

From the experimental point of view, the determination of the separation $|{\bi r}-{\bi q}|$, can also be obtained by the measurement of the internal angular velocity of the CC, since $R_0=c/\omega_0$. An experiment of the type of Gouan\'ere et al. \cite{Guanere} will ellucidate whether the internal frequency of the electron is the frequency postulated by De Broglie $\omega_0/2$, or Dirac's predicted frequency $\omega_0$, which is the frequency of this model as described in \ref{PlanckDB}. It has been suggested to enlarge the energy range of the above experiment \cite{clock}, to measure this internal frequency.

The integrations of this work have been performed numerically by means of the computer package {\it Dynamics Solver} \cite{JMA}. It uses a Dormand-Prince 8(5,3) integration code \cite{Dormand}. All codes have adaptive step size control and we check that smaller tolerances do not change the results. 
For the interested reader we include in the Appendix {\bf\ref{Append}} a link to two {\tt Mathematica} notebooks to perform all the above integrations with a widespread and different computer program.

The electric dipole moment is computed with 11 decimal digits while the magnetic moment with 12. More accuracy can be obtained by taking the time average over a greater number of turns. For instance, with $v=0.01$, of table {\bf{2}} when the spin and the velocity form an angle of $60^\circ$, the magnetic moment is $\mu=0.499956249432$, computed with $n=100$ turns, while for $n=300$ turns gives the value $\mu=0.4999562483785$, where the difference is at the 9th decimal position.
 
\section{Conclusions}
\label{conclusions}

Dirac's theory predicts the exact factor $g=2$ in the definition of the magnetic moment \cite{Dirac}. In a non-relativistic framework Levy-Leblond's equation \cite{LL1967} also predicts $g=2$, as well as in the classical model of the Dirac particle analyzed here \cite{Rivasbook}. However, when we try to measure the magnetic moment we have to interact with the electron. The interaction modifies the motion of the CC and CM of the Dirac particle and this justifies that the measurement of the magnetic moment must be different than the theoretical prediction for the electron at rest.

The conclusions about the measurements of the magnetic moment 
and the electric dipole moment of the electron are based on the analysis of
the classical electron model which leads to the Dirac equation when quantized, as described in Section {\bf\ref{clasicalelectron}}.

The main characteristic of this model is that the dynamics of the electron is the description 
of the motion of a single point ${\bi r}$, which is considered the center of charge of the electron. In addition, 
the electron has another characteristic point, the center of mass ${\bi q}$, which is always a different point from the center of charge.

If the electron has two characteristic points then it should be clear with respect to which point the angular momentum, that is, the spin, is defined. In Dirac's theory we have that Dirac's spinor depends only on a point ${\bi r}$, where the external fields are defined and therefore this point is interpreted as the center of charge of the electron. Dirac's spin is defined with respect to this point ${\bi r}$. 
This spin is never conserved. If we consider the angular momentum with 
respect to the center of mass it is a conserved angular momentum for the free particle and which precesses backwards with Larmor's angular velocity in a magnetic field. In the physics literature and in the mentioned experimental works there is no distinction between these two spins.

The magnetic moment is produced by the motion of the center of charge around the center of mass, also known in the physics literature as the {\it zitterbewegung}, which is a trajectory with curvature and torsion. The electric dipole moment with respect to the CM is due to the separation between these two points and not to an asymmetric distribution of the electron charge. It is not oriented in the direction of the spin but it is on a plane orthogonal to the spin and rotates with angular velocity $\omega_0$. Both momenta are defined with respect to the center of mass.

The internal motion of the CC at the speed of light is of such high frequency that from the classical point of view the measurement of the observables must be interpreted as a temporal average over a sufficiently long time. We have numerically solved the dynamical equations of the model in a Penning trap with the same characteristics and fields as in one of the experiments. In the classical description, we have assumed no thermal interaction and that the fields and surfaces of the cavity are those described theoretically. 

Measuring the electric dipole moment of the electron is to measure the average separation between the center of mass and the center of charge of the electron. In no case does this have anything to do with the spin precession angular velocity but depends on the velocity of the center of mass $v$ and the orientation $\alpha$ of this velocity with respect to the center of mass spin. This average value of the electric dipole moment is slightly larger than the theoretical value  predicted in Dirac's formalism and in our model. It is not an intrinsic property and its measurement depends on $(v/2)\sin\alpha$.

The conjecture that the electron's magnetic moment has the value $\mu=(g/2)\mu_B$, where $g\neq2$ is considered to be an intrinsic property, is a wrong conjecture since the measurement of the time-average value of the magnetic moment does not produce a single value but a value that is somewhat smaller than the theoretical prediction. It also depends 
on the value $(v/2)\sin\alpha$. It is clear that if we accept that the spin precession angular velocity is $\omega_s=g\omega_c/2$, the accurate measurement of these two angular velocities leads to a single result for the magnitude $g$. The assumption of formula (\ref{hipotesis}) about the
excitation energies of a spinning electron in a Penning trap is not justified on physical grounds, even if we consider that the spin is the CM spin.

The center of mass spin needs two complete cyclotron turns to return to its initial position because $\omega_s=-\omega_c/2$ (see figure {\bf\ref{fig:Rc5}}). This property is probably related to the geometric description of spinors which need a rotation of $4\pi$ to return to their initial position and which is related to the doubly connected structure of the rotation group.

The historical and critical analysis carried out by Consa \cite{Consa} on the theoretical and experimental measurement of the gyromagnetic ratio, as one of the greatest successes of Quantum Electrodynamics, casts doubt on the validity of both, the experiments and the theoretical analysis based on a theory full of divergences.
We believe in the accuracy of the experimental results but they are based on an erroneous theoretical interpretation of the magnitudes they attempt to measure. Their measurement of the spin precession angular velocity does not represent the measurement of the magnetic moment of the electron, because the assumption $\omega_s=g\omega_c/2$, is a wrong assumption.

If we take into account the efforts to accomodate the computed theoretical results with the experimental data of a magnitude like $g$, which is not an intrinsic property of the electron, makes us think that this is an inconsistent theoretical approach. 

\section{Appendix: Mathematica Notebooks}
\label{Append} 
\begin{figure}
\hspace{2cm}\includegraphics[scale=0.5]{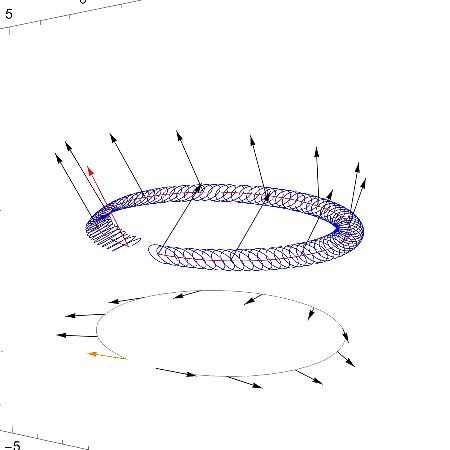}\includegraphics[scale=0.5]{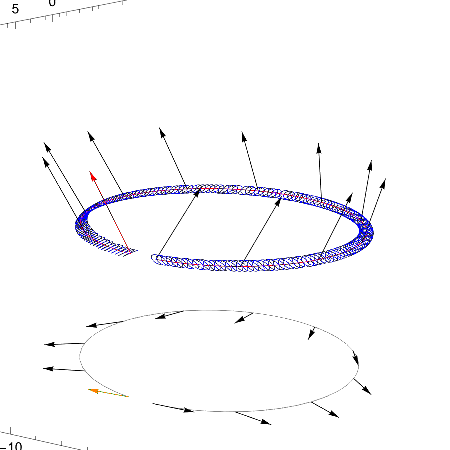}\includegraphics[scale=0.5]{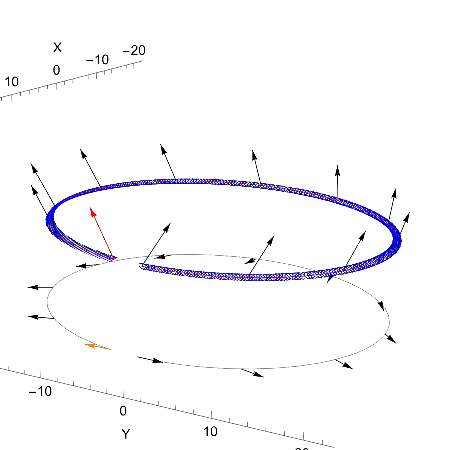}

\hspace{2cm}\includegraphics[scale=0.5]{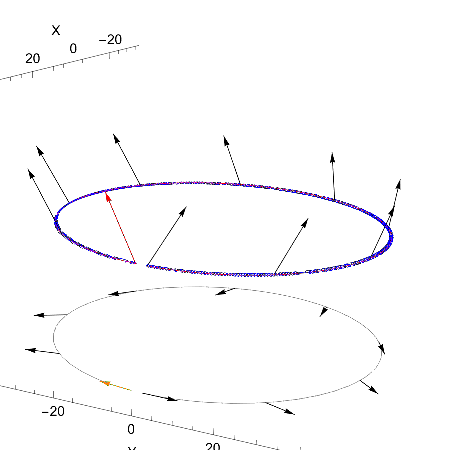}\includegraphics[scale=0.55]{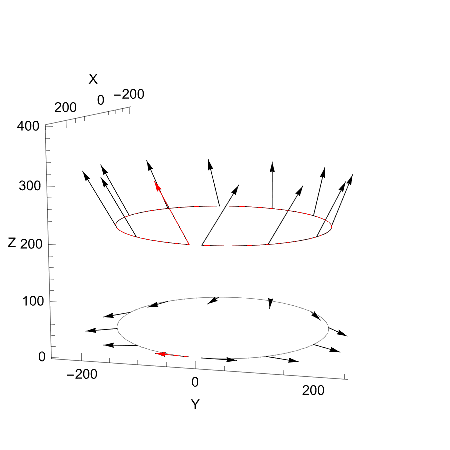}\includegraphics[scale=0.55]{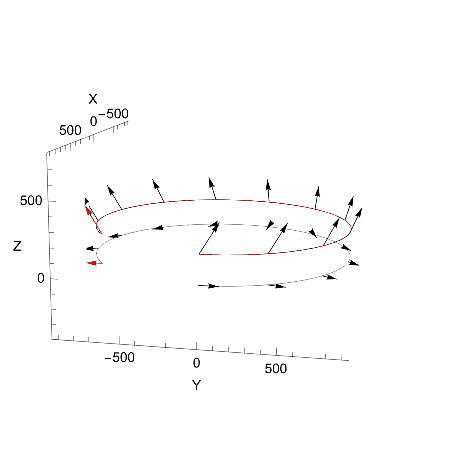}

\hspace{2cm}\includegraphics[scale=0.5]{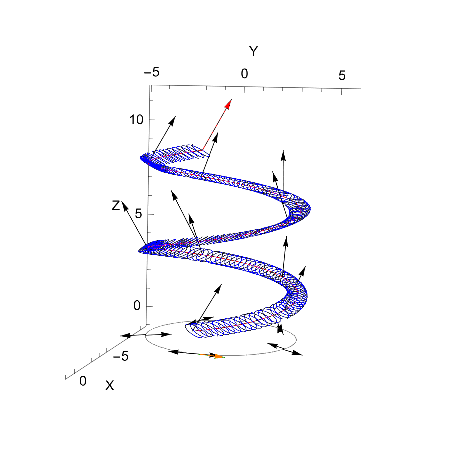}\includegraphics[scale=0.5]{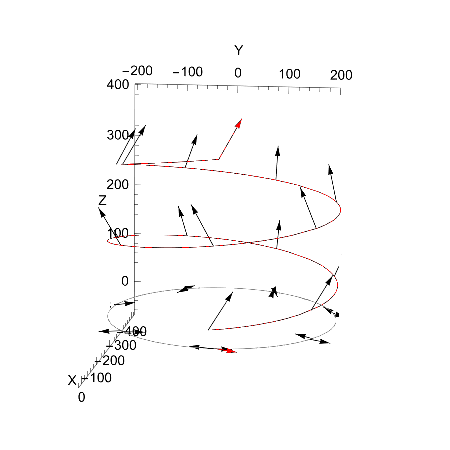}
\includegraphics[scale=0.7]{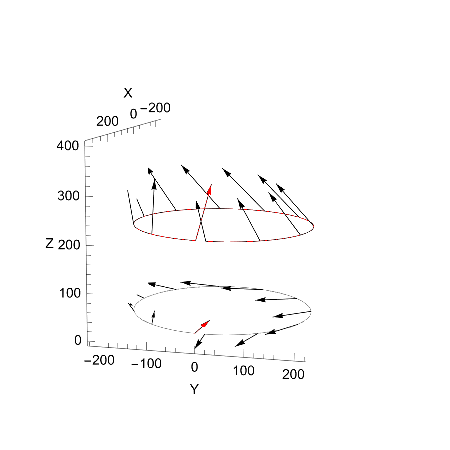}
\caption{Different cyclotron motions of radii $R_c=5,10,20,40,200$ and 800, with different values of the magnetic field and of the initial velocity of the CM. In all cases, except the last figure, the initial spin orientation is the same $\theta=30^\circ$, $\phi=90^\circ$, and always precesses backwards with Larmor's angular velocity. In the first four figures the motion of the CC is also visible at this scale but for larger cyclotron radii the motion of the CC is indistinguishable from the CM motion.\\ The first two figures of the third row show the two-turn cyclotron motion for two different radii with an initial component of the CM velocity along the magnetic field. After these two turns the CM spin returns to its initial orientation. The last figure in the third row is the same integration as the central figure but with a different orientation of the initial spin. After almost one turn the spin projection is opposite to the initial spin projection which justifies that the spin precesses backwards with $\omega_c/2$. The interested reader can modify the fields, the boundary conditions, the spin orientations and rotate the 3D figures to better appreciate the zitterbewegung and the cyclotron motion. The zitter motion of the CM is not noticeable at this scale, and the spin $S=1/2$ has been rescaled in every picture.}  
\label{fig:Rc5}
\end{figure}

Numerical integrations have been performed with the use of a computer program \cite{JMA}. It is widespread among physicists the use of the {\tt Mathematica} \cite{Math} computer program for solving analytical and numerical problems. With the help of Juan Barandiaran we include in this Appendix a link to two Mathematica notebooks \cite{Notebooks} {\tt PointParticleinPenningTrap.nb} and {\tt DiracParticleinMagneticField.nb} with which the interested reader can perform different calculations, such as the motion of the spinless particle in a Penning trap of the figure {\bf{2}} and its analytical solution, and that of the Dirac Particle in a uniform magnetic field of the figures {\bf\ref{fig:cyclo01}},  {\bf\ref{fig:spindynamics}}, {\bf\ref{fig:spin100}} and {\bf\ref{fig:Rc5}}. 

In these notebooks, which can be obtained through the aforementioned Zitter Institute website, you can change the fields, the spin orientation and the electron boundary conditions to your liking in order to analyze the behavior of the classical Dirac particle in different situations. The pictures are three-dimensional and can be rotated as you like to appreciate the zitterbewegung and the spin orientations. We draw the trajectories of the CC and CM and also their projections on the $XOY$ plane, the spin ${\bi S}_{CM}$ and its projection on the $XOY$ plane at different CM positions at the figure {\bf\ref{fig:Rc5}}, for cyclotron motions of different radii $R_c$, where the last spin is represented in red colour and for different values of the magnetic field and of the initial velocity of the CM. The initial spin in all integrations, except in the last picture, has the same orientation. The important feature is that the spin always precesses backwards with half the cyclotron angular velocity $\omega_s=-\omega_c/2$, irrespective of the particle velocity and external magnetic field, because Dirac's spin ${\bi S}$ satisfies the dynamical equation $d{\bi S}/dt={\bi p}\times{\bi u}$. All these integrations are included as bookmarks of the notebook {\tt DiracParticleinMagneticField.nb}. 

One turn of the CC takes place in a time $\pi$ in dimensionless units which corresponds to $\simeq10^{-21}$s and is also the integration time variable. 
{\tt Mathematica}, when computing the integration of a trajectory, keeps in memory all data until the end of integration to depict the solution. For very long integration times aborts the calculation.
In a magnetic field of 5 Teslas with a velocity $v/c=0.1$, the cyclotron radius is $R_c=3.4\cdot{10^{-5}}$m which in dimensionless units corresponds to a radius of $\widetilde{R}_c=3.53\cdot 10^8$, and to depict one cyclotron trajectory requires an integration time of  $2.1\cdot 10^{10}$. A personal computer has no sufficient memory to record so many data and the integration program cancels. For longer integration times a big computer is required.

\ack{I am very grateful to my colleague Juan M Aguirregabiria for the use of his computing program DS-Solver \cite{JMA} with which numerical calculations and figures have been made.\\
I am greatly indebted to Juan Barandiaran for his help and for preparing the two {\tt  Mathematica} notebooks mentioned in the Appendix \ref{Append}}.

\end{document}